\documentclass[fleqn,usenatbib]{mnras}

\usepackage{newtxtext,newtxmath}

\usepackage[T1]{fontenc}
\usepackage{ae,aecompl}

\DeclareRobustCommand{\VAN}[3]{#2}
\let\VANthebibliography\thebibliography
\def\thebibliography{\DeclareRobustCommand{\VAN}[3]{##3}\VANthebibliography}

\usepackage{graphicx}	
\usepackage{amsmath}	
\usepackage{subcaption}
\usepackage{color}
\usepackage{ulem}
\usepackage{hyperref}
\usepackage{soul}
\usepackage{longtable}
\usepackage{scrextend}
\usepackage{booktabs}

\renewcommand{\Vec}[1]{\bmath{#1}}
\newcommand{\grale}{\textsc{grale}}
\newcommand{\lenstool}{\textsc{Lenstool }}
\newcommand{\hst}{{\it HST}}

\newcommand{\Review}{\textcolor[rgb]{0,0,0}}

\title[{\sc Grale} inversion of A1689]{An excursion into the core of the cluster lens Abell 1689
}

\author[A. Ghosh et al.]{Agniva Ghosh,$^{1}$\thanks{e-mail:  ghosh116@umn.edu} 
Dominic Adams,$^{1}$ 
Liliya L.R. Williams,$^{1}$\thanks{e-mail: llrw@umn.edu}  
Jori Liesenborgs,$^{2}$
Anahita Alavi$^{3}$ and
\newauthor
Claudia Scarlata$^{1}$\\
\\
$^{1}$School of Physics and Astronomy, University of Minnesota, 116 Church Street SE, Minneapolis, MN 55455, USA\\
$^{2}$UHasselt - tUL - Flanders Make, Expertisecentrum voor Digitale Media, Wetenschapspark 2, B-3590, Diepenbeek, Belgium\\
$^{3}$IPAC, California Institute of Technology, 1200 E. California Boulevard, Pasadena, CA 91125, USA
}

\date{Accepted XXX. Received YYY; in original form ZZZ}

\pubyear{2021}

\begin{document}

\label{firstpage}
\pagerange{\pageref{firstpage}--\pageref{lastpage}}
\maketitle

\begin{abstract}

Abell 1689 is a well studied cluster of galaxies and one of the largest gravitational lens systems ever observed. We have obtained a reconstruction of the cluster Abell 1689 using \grale, a free-form lens inversion method that relies exclusively on the multiple image data. Non-inclusion of any data related to cluster member galaxies ensures an unbiased measure of the mass distribution, which is the most notable feature of free-form methods like \grale. We used two different sets of multiple image systems from the available strong lensing data - one containing only the secure systems (107 images), and the other containing all available systems, only excluding some very non-secure systems (151 images). 
For the very well-constrained central $\sim$100 kpc region of the cluster we made detailed comparison of the \textsc{grale} reconstructed lensing mass and stellar mass retrieved by the Spectral Energy Distribution (SED) fitting software \textsc{fast++}. 
We found a light-unaccompanied mass peak in this region, whose {existence, while tentative, is favored by the distribution of nearby images that are local maxima in the Fermat potential.} {However, further tests, using different methodologies are needed to confirm the reality of this feature. If it shown to be real, this} light-unaccompanied mass peak is consistent with dark matter self-interaction cross-section $\sigma \lesssim 1$cm$^2$/g, while 
being in tension with larger cross-sections.

\end{abstract}




\section{Introduction}

Galaxy clusters are the most recently formed gravitationally bound structures in the hierarchical structure formation in $\Lambda$CDM (Lambda Cold Dark Matter)  model of cosmology.  Galaxy clusters contain hundreds of galaxies and X-ray emitting hot intracluster plasma, but their mass budget is dominated by dark matter. The mass distribution within a galaxy cluster can be measured through several physical processes such as kinematics of member galaxies \citep[e.g.,][]{Saro2013,rin22}, the hydrodynamics of the hot gas \citep[e.g.,][]{Vikhlinin2009,Ettori2013}, and gravitational lensing.

The strong lensing effect is most powerful in the central regions $\lesssim 0.3\,$Mpc of galaxy clusters, which contain high concentrations of both baryonic and dark matter, and often produce high density of multiple images. These central regions serve as cosmic laboratories for studying the interplay between baryonic and dark matter. In the current $\Lambda$CDM model of cosmology, dark matter is assumed to be cold and collisionless. However, if dark matter exhibits self-interaction it is highly likely that the resulting effects will be exhibited in the central dense regions of clusters \citep{Kneib2011}. For example, self-interaction can result in offsets between galaxies and their dark matter halos \citep{Harvey19}, or erase small scale mass features, like central density cusps. In recent times, strong gravitational lensing by galaxy clusters has been extraordinarily useful for the study of the high redshift universe \citep{Lotz2017,Bouwens2017a,Livermore2017,Ishigaki2018}, as clusters can magnify faint distant background sources with fixed surface brightness, by increasing their angular extent, and hence observed fluxes, thus pushing the sources above the detection threshold of contemporary telescopes \citep{Schneider1992,Bartelmann2010}.  Multiple images produced through strong lensing by individual cluster member galaxies are also currently detectable in the deepest {\it Hubble Space Telescope} (\hst) observations \citep{Meneghetti2020}.

These studies rely on the mass distributions within the clusters, which are made possible by lens reconstruction algorithms. Several lens reconstruction techniques have been developed and widely tested within the lensing community  over the last few decades \citep[see for example,][]{Meneghetti2017}. These techniques predominantly use the strongly lensed multiple images, but some also incorporate weak lensing data in their analysis. The number of observed strongly lensed multiple images in a galaxy cluster depends on the depth of the observations, the source number density behind the cluster, \Review{mass of the cluster} and the cluster redshift. The maximum number of observed strongly lensed images in a galaxy cluster is currently in the order of $\sim\!100$, which is insufficient to break all possible lensing degeneracies \citep{Liesenborgs2012, Ghosh2020}. This leads to different reconstruction techniques producing non-unique solutions \citep{Limousin2016,Priewe2017, Meneghetti2017}. However, with the inception of the {\it Hubble} Frontier Fields survey (HFF; PI: J. Lotz) with the \hst, it has become possible to estimate systematic uncertainties based on different reconstructions of the same clusters of galaxies \citep{Lotz2017}. 

Reconstruction techniques can be broadly classified into two categories: simply parametrized and free-form. Simply-parametrized methods, which are widely used, assign simple mass profiles to cluster galaxies, like Navarro-Frenk-White \citep[NFW;][]{Navarro1996}, Pseudo Isothermal Elliptical Mass Distributions \citep[PIEMD; ][]{Kovner1993}, pseudo-Jaffe distributions \citep[PJ;][]{Keeton2001}.
The cluster-scale dark matter distribution is represented by the same or similar profiles, but with larger scale lengths. \Review{Parametric methods are physically motivated and are especially helpful when the number of input lensed images is small.}

In contrast to parametric models, free-form methods, such as \textsc{grale} \citep{Liesenborgs2006a,Liesenborgs2020}, do not assume any relation between the distribution of mass and light.  
The advantage of free-form methods is that they use the lensed images only, avoiding the use of strong priors of simply-parametrized methods.
While simply-parametrized priors are astrophysically motivated, they describe the average properties of galaxies and clusters, and may not correctly capture the mass distribution, especially in merging and non fully equilibrium clusters. Because of their high sensitivity to lensed images, free-form methods can detect cluster mass features that elude parametric methods \citep{Ghosh2021}. \Review{However, in the case of \grale, which constructs a large number of mass models, flexibility can lead to unastrophysical models that are also included in the ensemble average maps.}

Abell 1689 (hereafter A1689) is a well studied cluster of galaxies and one of the largest gravitational lens systems ever observed, owing to its large mass. It is one of the earliest clusters to be revealed with more than $\sim\!100$ multiple images from 30 sources \citep{Broadhurst2005}. This was a giant leap forward in the number of strong lensing constraints for a given cluster in pre-HFF era. A1689 has been modelled by a number of mass modeling groups using parametric methods \citep{Broadhurst2005, Zesker2006, Halkola2006, Limousin2007}, free-form methods \citep{okura08,Coe2010,Mohammed2014,Cha2022}, and hybrid methods \citep{Diego2005,Diego2015} using only strong lensing data, or a combination of both strong and weak data. The number and quality of multiple images have increased with the discovery of new systems and updates to the existing systems with spectroscopic redshifts \citep{Limousin2007,Coe2010,Diego2015,Bina2016,Alavi2016}. 

\Review{A1689 appears roughly circularly symmetric on the sky, and is likely in hydrostatic equilibrium with a highly symmetric X-ray morphology \citep{Lemze2008,Lemze2009,riemer09}.} It is a cool core cluster, with a reasonably good agreement between its weak lensing mass and X-ray mass within $\sim 0.9$Mpc \citep{riemer09}. The mass profile of A1689 agrees well with the standard Navarro, Frenk and White (NFW) fitting formula profile, consistent with the $\Lambda$CDM cosmology but with a high concentration \citep[$c=6-11$; see table 2 of][]{Coe2010}. \citet{Duffy2008} estimated the NFW concentration value for clusters of similar mass from three N-body simulations to be $c = 3-4$ which is roughly the same as estimated by \cite{Umetsu17} from observations of a large sample of clusters of similar mass. \citet{Peng2009} combined weak-lensing and X-ray data from the Chandra satellite and found a discrepancy between the hydrostatic mass from X-ray data and lensing masses. They attributed this discrepancy to  projection effects. {\cite{mor11a} modeled the cluster as a triaxial mass distribution, finding that this resolved the discrepancies found earlier, and resulted in the dark matter density profile slope that was consistent with $\Lambda$CDM predictions. A further analysis by \cite{mor11b} added weak lensing data, and placed constraints on the amount of non-thermal gas pressure. In agreement with the earlier dark matter halo shape conclusions,} \citet{Sereno2013}, using X-ray and Sunyaev–Zel'dovich (SZ) measurements found that the cluster has a triaxial shape elongated along the line of sight. The multi-probe analysis by \cite{Umetsu2015} using strong and weak lensing along with X-ray and SZ data also favored a triaxial geometry for A1689.

In this paper we focus on two aspects of our reconstructed mass distribution of A1689, that relate to the properties of dark matter: the global, radially averaged density profile of the cluster, and the detailed comparison between the stellar light and mass density vs. the total density in the central $\sim$100 kpc region, that contain 4 massive ellipticals. 
The comparison between stellar mass and total mass can reveal particle nature of dark matter, specifically, its self-interaction cross section.

The paper is structured as follows: In Section~\ref{sec:data}, we provide details of the HST imaging and the observed strong lensed multiple images for A1689. In Section~\ref{sec:reconstruction}, we discuss the details of the reconstruction algorithm \textsc{grale} and the results obtained from the mass reconstructions. In Section~\ref{sec:fast}, we discuss the Spectral Energy Distribution (SED) fitting software \textsc{fast++} which we use to obtain the stellar mass distribution in the central $\sim\!100$ kpc region of the cluster, and in Section~\ref{sec:sedresults} we compare between the stellar mass and the lensing mass. Finally, we draw our conclusions with further discussions on the results in Section~\ref{sec:discussion}. \Review{To test the impact of the resolution of our lens reconstruction method on the main conclusions, in Appendix~\ref{sec:lores} we present results using lower resolution inversions.}

Throughout this work we used the $\Lambda$CDM model of cosmology: flat, with matter density, $\Omega_m = 0.3$, cosmological constant density, $\Omega_{\Lambda}= 0.7$, and the dimensionless Hubble constant $h = 0.7$. The redshift of A1689 is, $z_l=0.1832$.  The center of the reconstruction region is at R.A.$=197.873^{\circ}$, Dec.$=-1.338^{\circ}$. At the redshift of the cluster, 1 arcsec corresponds to $3.08$  kpc.

\section{Data}
\label{sec:data}
\subsection{HST Data}
\label{sec:hstdata}
In this paper, we use public imaging data from the Hubble Space Telescope (HST). A1689 has been observed with Advanced Camera for Surveys Wide Field Channel \citep[ACS/WFC;][]{acs2022} in four optical bandpasses (F475W, F625W, F775W, and F850LP) as part of the HST program ID 9289 (PI: H. Ford), taken in Cycle~11 in June 2002 and in the F814W optical bandpass as part of the HST program ID 11710 (PI: J. Blakeslee), taken in Cycle~17 in May to July 2010; and in three Wide Field Camera 3 Ultraviolet and VISible light \citep[WFC3/UVIS;][]{wfc2022} bandpasses (F225W, F275W, and F336W) as part of program IDs 12201 and 12931 (PI: B. Siana), taken in Cycle 18 in December 2010 and Cycle 20 in February and March 2012, respectively. More details on the data calibration and reduction can be found in \citet{Alavi2014,Alavi2016}. The images are all mapped to the same pixel scale of 80 mas pixel$^{-1}$.

\subsection{Lensing Data}
\label{sec:lensingdata}
\begin{figure*}
	\includegraphics[width=0.7\textwidth]{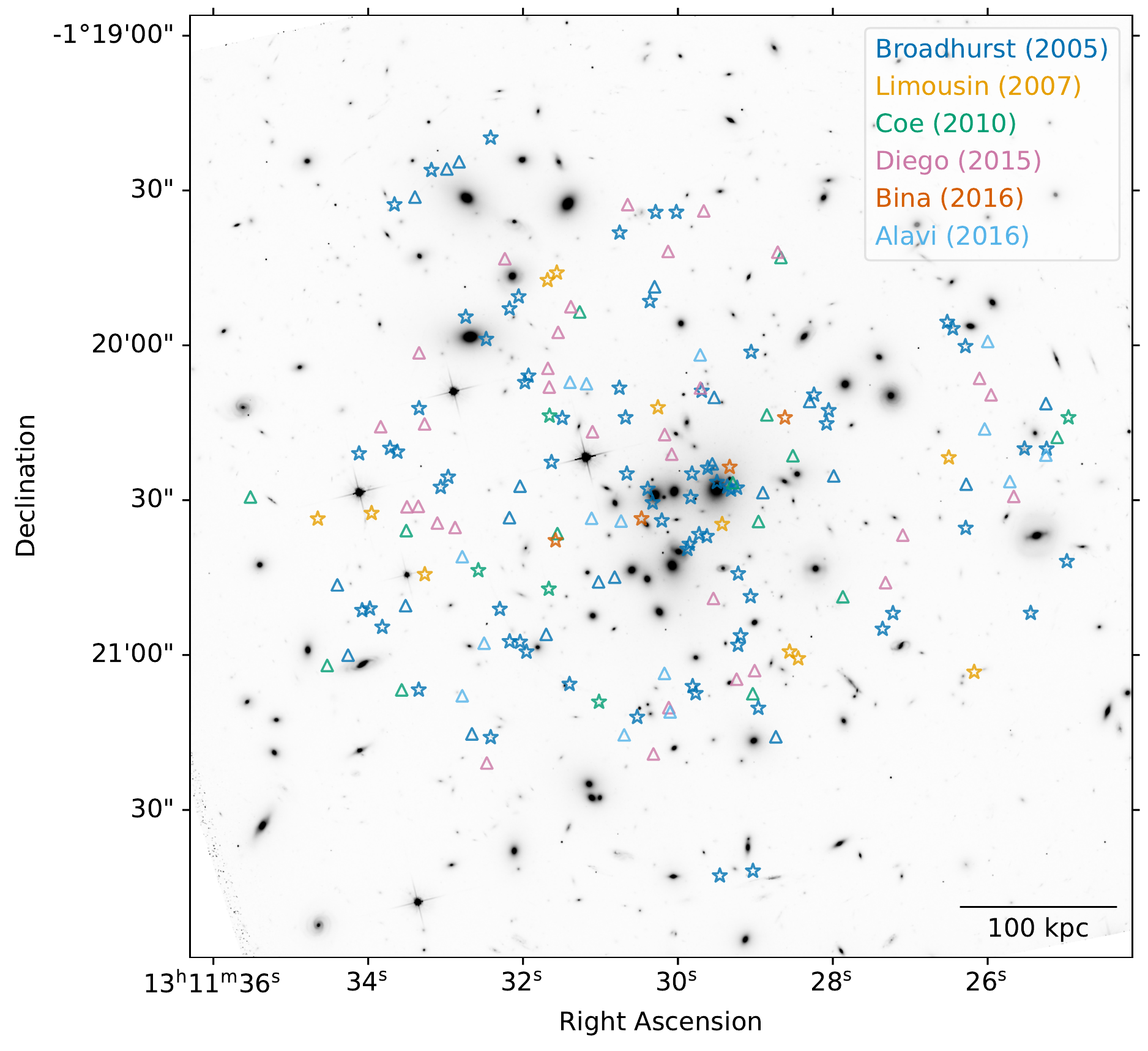}
    \caption{Strongly lensed images observed in A1689. Different colors show the images identified by various previous works to date, as mentioned in the plot legend. Star symbols show the systems with spectroscopic redshifts, and triangles denote systems with photometric redshifts. The background mosaic is in ACS/F814W filter. }
    \label{fig:images}
\end{figure*}

Strongly lensed multiply imaged systems in A1689 can be found in the existing literature \citep{Broadhurst2005,Halkola2006,Limousin2007,Coe2010,Diego2015}. A compilation of the lensed images is provided in Table~\ref{tab:imgdata}.  For system identification we maintained the notation used in \citet{Diego2015} (see their appendix A). In addition to the images listed in \citet{Diego2015,Diego2018cat} we have also added two new multiply imaged systems identified by \citet{Bina2016} as system 62 and 63, and five new multiply imaged systems identified by \citet{Alavi2016} as systems $64-68$ in  Table~\ref{tab:imgdata}, for completeness. Systems $1,4,7,8,17,46$ and $50$ are also updated with spectroscopic redshifts from the {Multi-Unit Spectroscopic Explorer} \citep[MUSE;][]{Bina2016}. These observed strongly lensed multiple images are shown in Figure~\ref{fig:images} overlaid on the ACS/F814W mosaic.

\section{Lens Reconstruction}
\label{sec:reconstruction}
\subsection{Method: Strong Lens Reconstruction Using \textsc{grale}}
\label{sec:grale}
The lens inversion method used in this paper is based on the reconstruction code \textsc{grale}\footnote{\textsc{grale} is publicly available. For the software and tutorials visit \url{https://research.edm.uhasselt.be/jori/grale2/}}. The publicly available {\grale}  software implements a flexible, free-form, adaptive grid lens inversion method, based on a multi-objective genetic algorithm (GA) as the underlying optimization procedure \citep{Liesenborgs2006a,Liesenborgs2007,Liesenborgs2020}. \textsc{grale} uses only multiple lensed image locations and their redshifts as input.  It is ideally suited for reconstructions with numerous multiple images, available with \hst~data.  The fact that the number of its model parameters exceeds the number of data constraints allows a fuller exploration of degenerate mass distributions
\citep{Mohammed2014,sebesta2016,sebesta2019,Meneghetti2017,williams2018,williams2019,Ghosh2020,Ghosh2021}. We refer the readers to see section 3.1 of \citet{Ghosh2020}, for a concise description of the modus operandi of \textsc{grale}. In short, \grale's mass basis set consists of hundreds of projected Plummer spheres, each occupying a grid cell in the lens plane. A run starts with a coarse uniform grid made up of a small number of grid cells of equal size. As the code runs the grid is refined, especially in areas of high density. At each iteration, the code optimizes the Plummer masses to produce overlapping images when projected back onto the source plane. The final map has a few thousand grid cells ranging from 0.66 to 42.5 arcsec, each occupied by a Plummer proportional to the cell width, and mass determined by the genetic algorithm. The angular resolution of the reconstruction is set by the smallest Plummer.

\Review{To test the effect of the resolution we run two sets of reconstructions. The main paper contains high resolution results, while Appendix~\ref{sec:lores} presents results using lower resolution runs, which is typically 0.5 times lower than those in the main paper, and is similar to those in earlier papers that use \grale.}

\begin{figure*}
     \centering
     \begin{subfigure}[b]{0.49\textwidth}
         \centering
         \includegraphics[width=\textwidth]{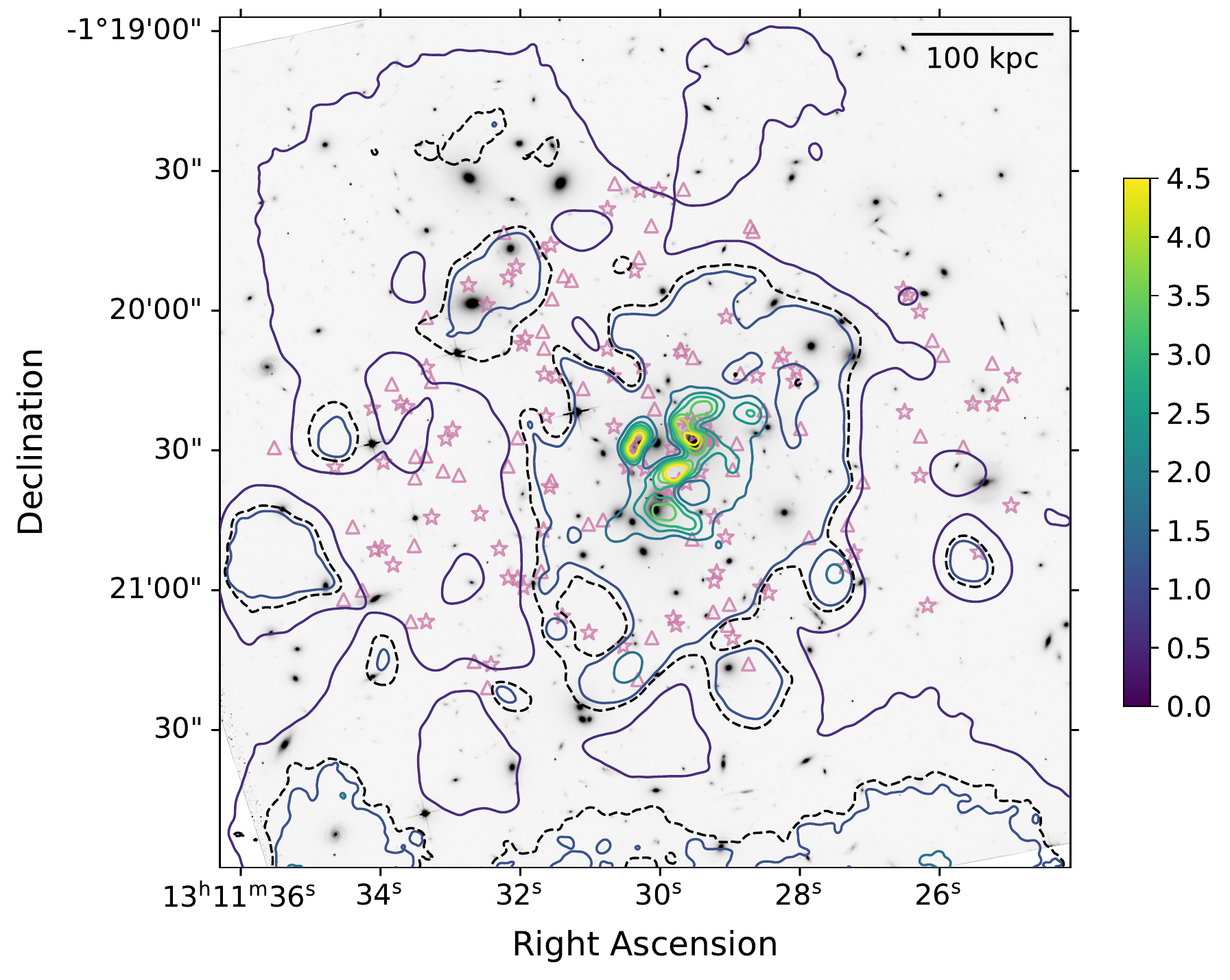}
         \caption{ALL-reconstruction}
     \end{subfigure}
     \quad
     \begin{subfigure}[b]{0.49\textwidth}
         \centering
         \includegraphics[width=\textwidth]{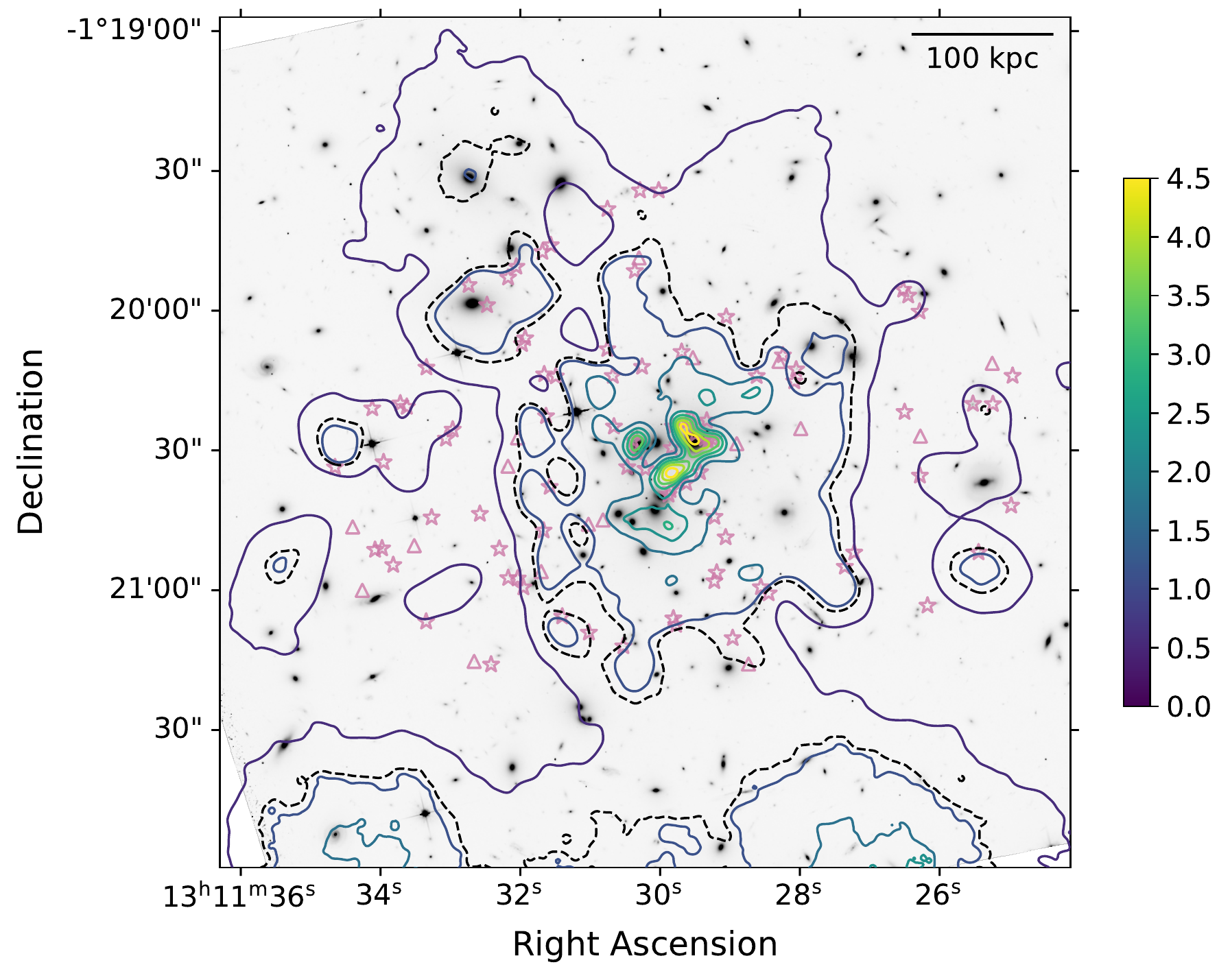}
         \caption{S-reconstruction}
     \end{subfigure}
     
     \caption{The coloured contours of \textsc{grale} reconstructed projected surface mass density distributions of A1689 ($z_l=0.1832$) overlaid on the mosaic of A1689 in the ACS/F814W filter. The surface mass density values are scaled by $\Sigma_{\rm{crit,0}}=5.47$ kg/m$^2$. At $z_l=0.1832$, 1 arcsec corresponds to approximately $ 3.082\,$kpc. The black dashed contour corresponds to $\kappa=1$, i.e. the contour of $\Sigma_{\rm{crit,0}}$.  Purple stars and triangles shows the multiple lensed images used to perform the reconstructions with spectroscopic and photometric redshifts, respectively.}
    \label{fig:massmaps}
\end{figure*}

\begin{figure*}
     \centering

     \begin{subfigure}[b]{0.49\textwidth}
         \centering
         \includegraphics[width=\textwidth]{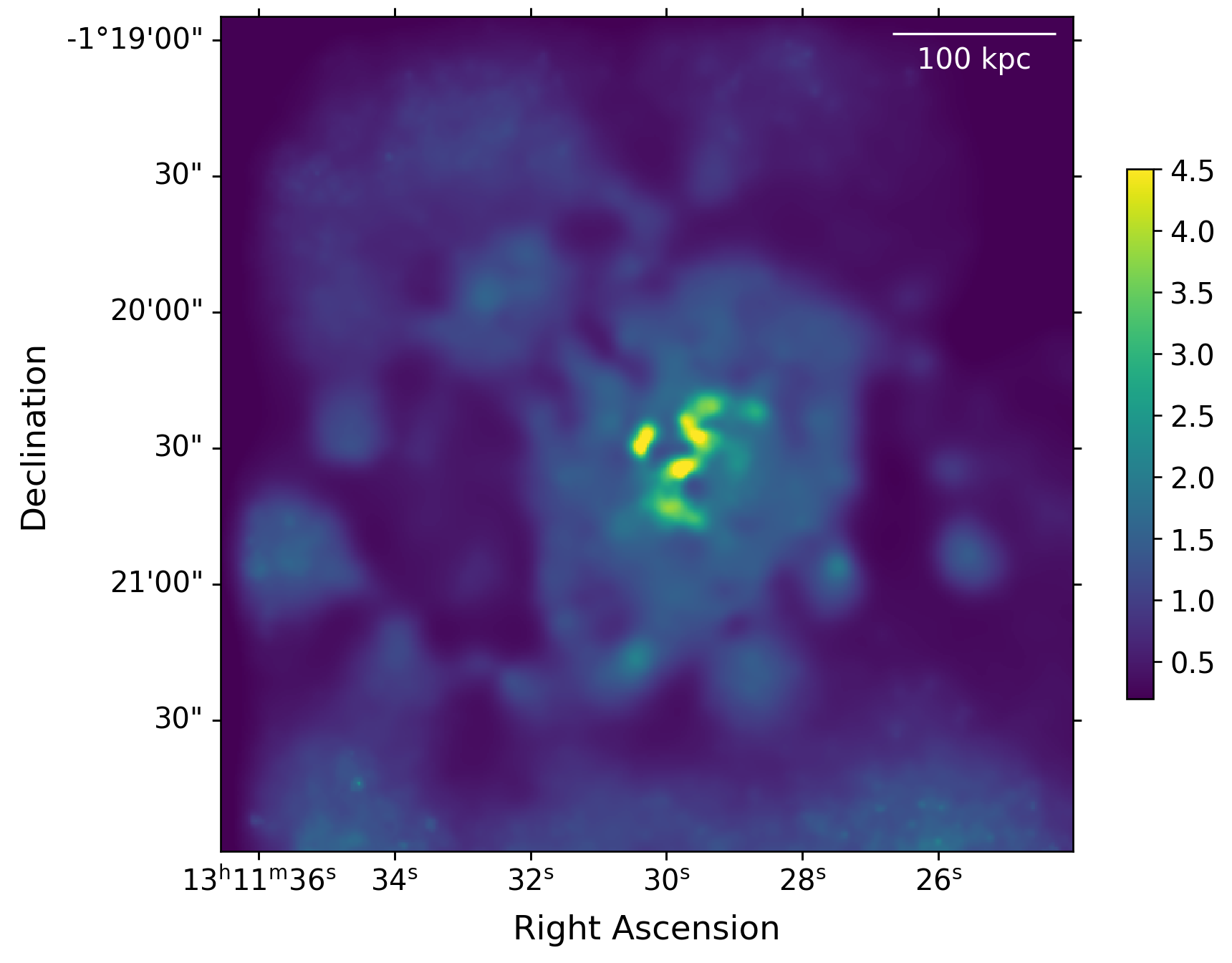}
         \caption{ALL-reconstruction}
     \end{subfigure}
     \quad
     \begin{subfigure}[b]{0.49\textwidth}
         \centering
         \includegraphics[width=\textwidth]{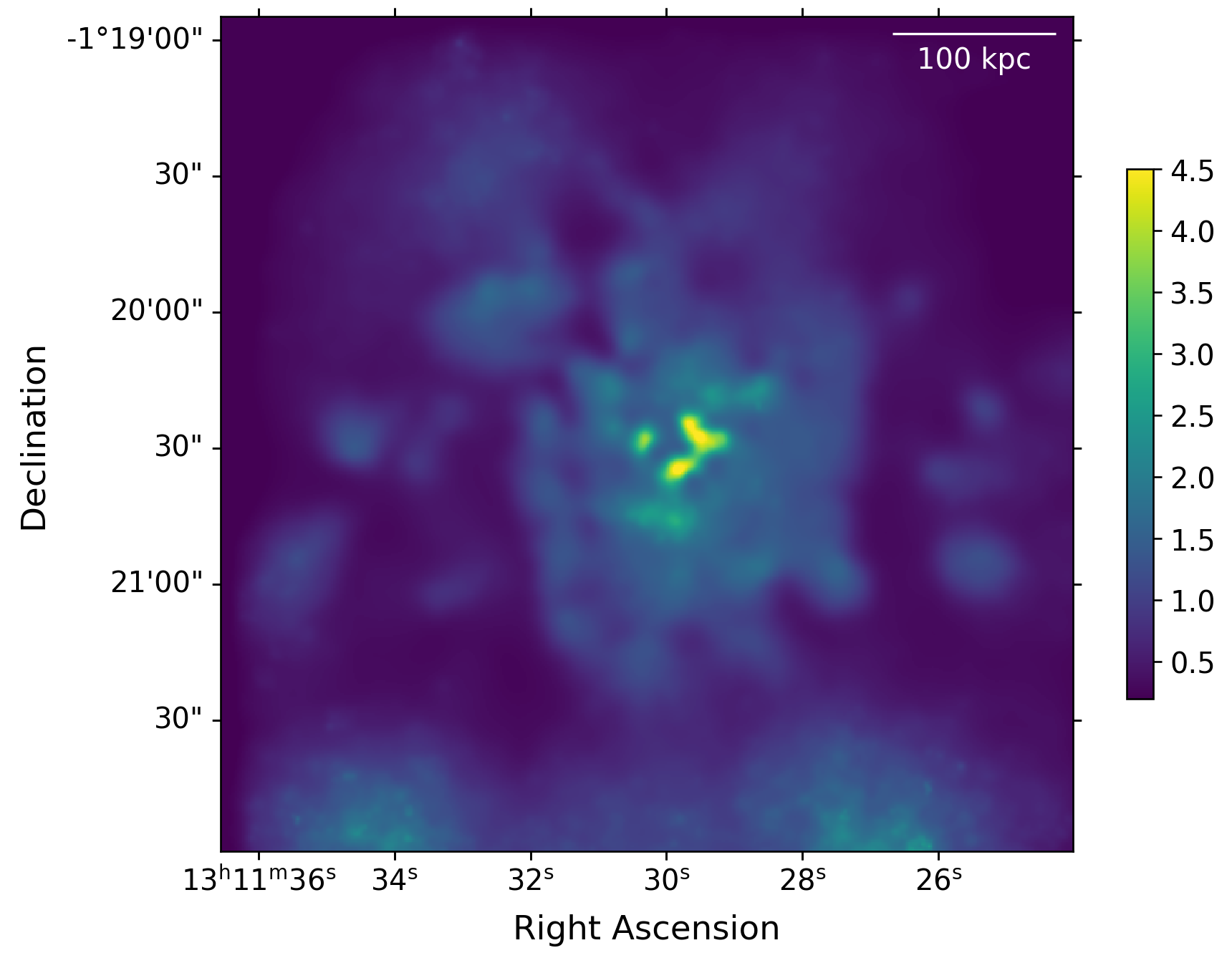}
         \caption{S-reconstruction}
     \end{subfigure}
     \caption{Same as Figure~\ref{fig:massmaps} but with full colormaps for the reconstructed mass distributions.}
    \label{fig:massmaps_fc}
\end{figure*}

\begin{figure*}
     \centering

     
     \begin{subfigure}[b]{0.475\textwidth}
         \centering
         \includegraphics[width=\textwidth]{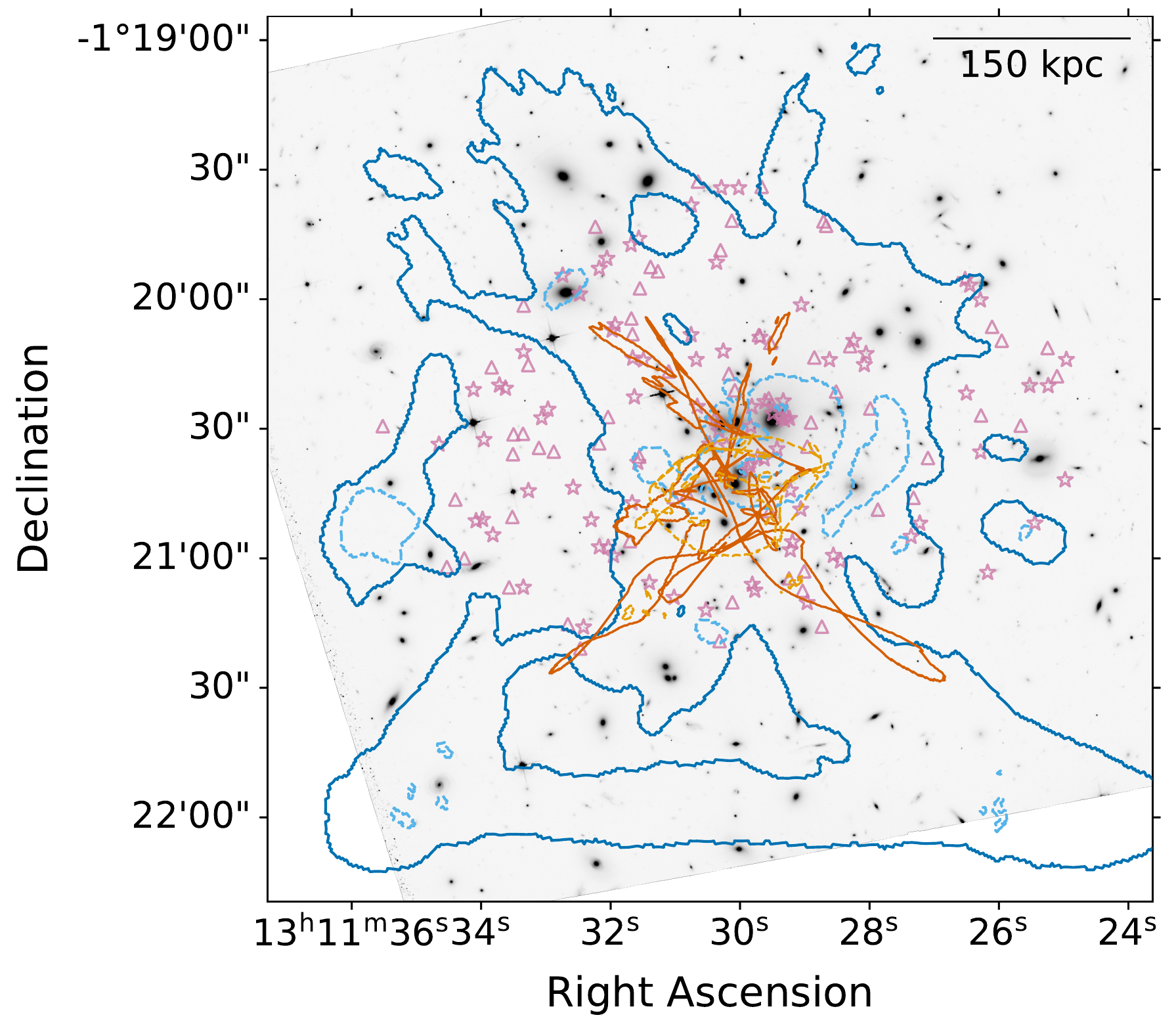}
         \caption{ALL-reconstruction}
     \end{subfigure}
     \quad
     \begin{subfigure}[b]{0.475\textwidth}
         \centering
         \includegraphics[width=\textwidth]{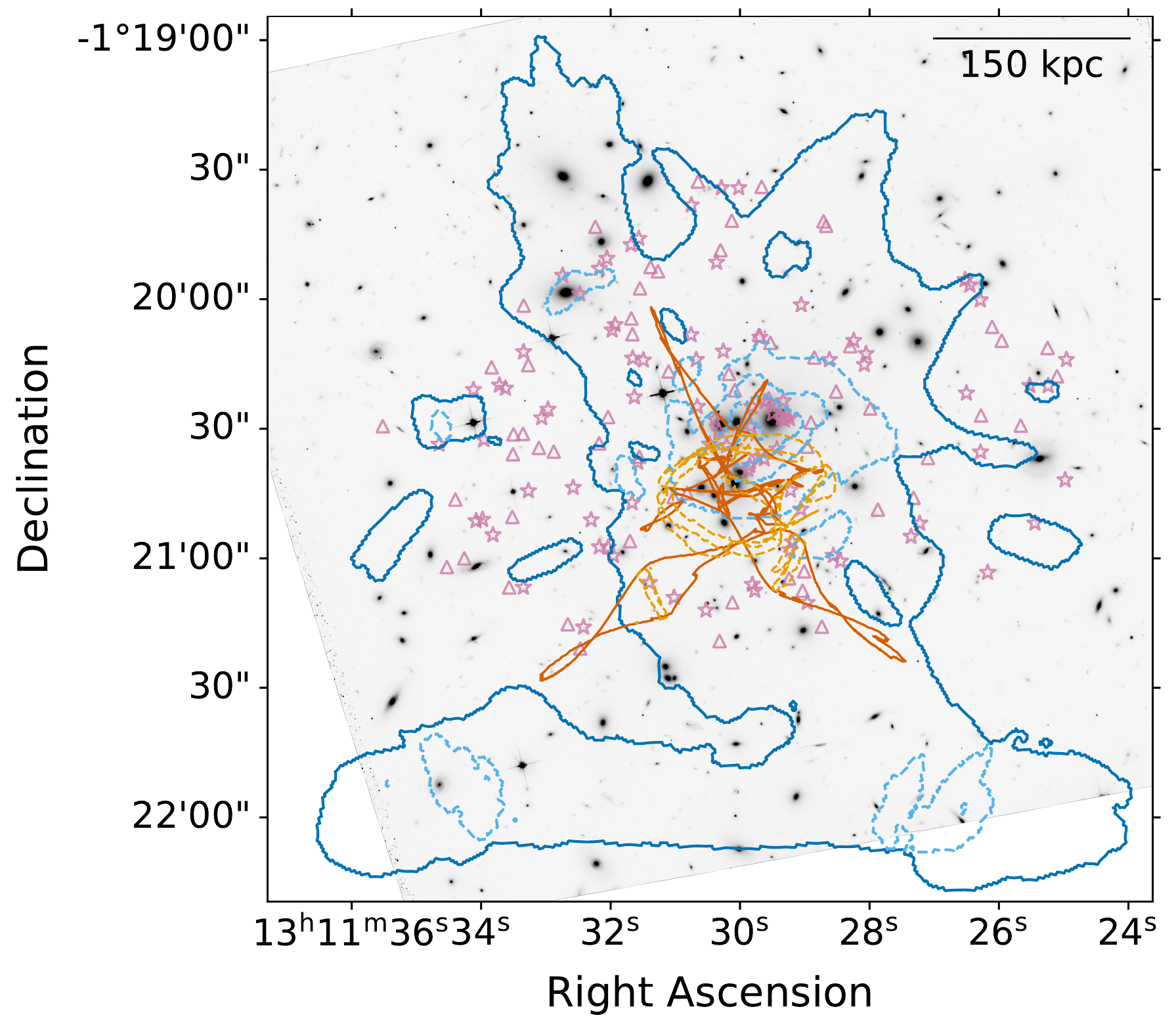}
         \caption{S-reconstruction}
     \end{subfigure}
     
        \caption{Critical lines and caustics for A1689 for source redshift of $z_s=9.0$ for the \textsc{grale} reconstructions overlaid on the mosaic of A1689 in the ACS/F814W filter. The solid and dashed blue contours represent the outer and inner critical curves, whereas the solid and dashed orange contours represent the tangential and radial caustics, respectively. Purple stars and triangles shows the multiple lensed images used to perform the reconstructions with spectroscopic and photometric redshifts respectively.}
        \label{fig:magmaps}
\end{figure*}

\subsection{Input: Multiple Image Data}
\label{sec:input}
Many multiple images are available for A1689, but not all are of the same quality. For example, spectroscopic redshifts are more precise and accurate compared to photometric ones. Since \textsc{grale} uses only the strongly lensed multiple image data, with no priors about the cluster member galaxies, it is particularly sensitive to the quality of the lensed image data. Therefore we select lensed images of higher quality.

As mentioned by \citet{Diego2015} systems $25$ and $32$ contain multiple candidates for some of the  counter-images, and systems $20, 25, 26$ and $27$ have shown different colors between different counter-images with the inclusion of the IR data by \citet{Diego2015}; these systems thus were not used in our work. We used the updated re-matching of the systems 10 and 12 by \citet{Diego2015} based on IR data (see their section 3 for more details about these systems).
We have also excluded the three systems 13, 14 and 30: since they do not have counterparts on the opposite side of the cluster, they can only constrain mass in their small immediate vicinity, and are therefore not good candidates for \grale.

For the reconstructions performed in this work we have used two different sets of input image data. The first set consists of the secure\footnote{Morphologically similar system of lensed multiple images with consistent colors, and redshifts.} systems (the `S-reconstruction') excluding all the systems excluded above and the not secure systems with photometric redshifts ($41-44,48,49,51,61$ and $64-68$) identified by \citet{Diego2015,Coe2010} and \citet{Alavi2016}. There are a total of 107 images from 31 sources in this set - 24 of these systems have spectroscopic redshifts and 7 have photometric redshifts. The second set includes all the available images except the already excluded systems, and the systems $64-68$ identified by \citet{Alavi2016}, (the `ALL-reconstruction') since they are not very secure because they have inconsistent photometric redshifts, whereas \textsc{grale} requires all lensed images from the same source to have the same redshift. There are a total of 151 images from 47 sources in this set - 24 of these systems have spectroscopic redshifts and 23 have photometric redshifts.

\subsection{Results: Mass Reconstruction}
\label{sec:reconstructionresults}

\begin{figure}
	\includegraphics[width=\columnwidth]{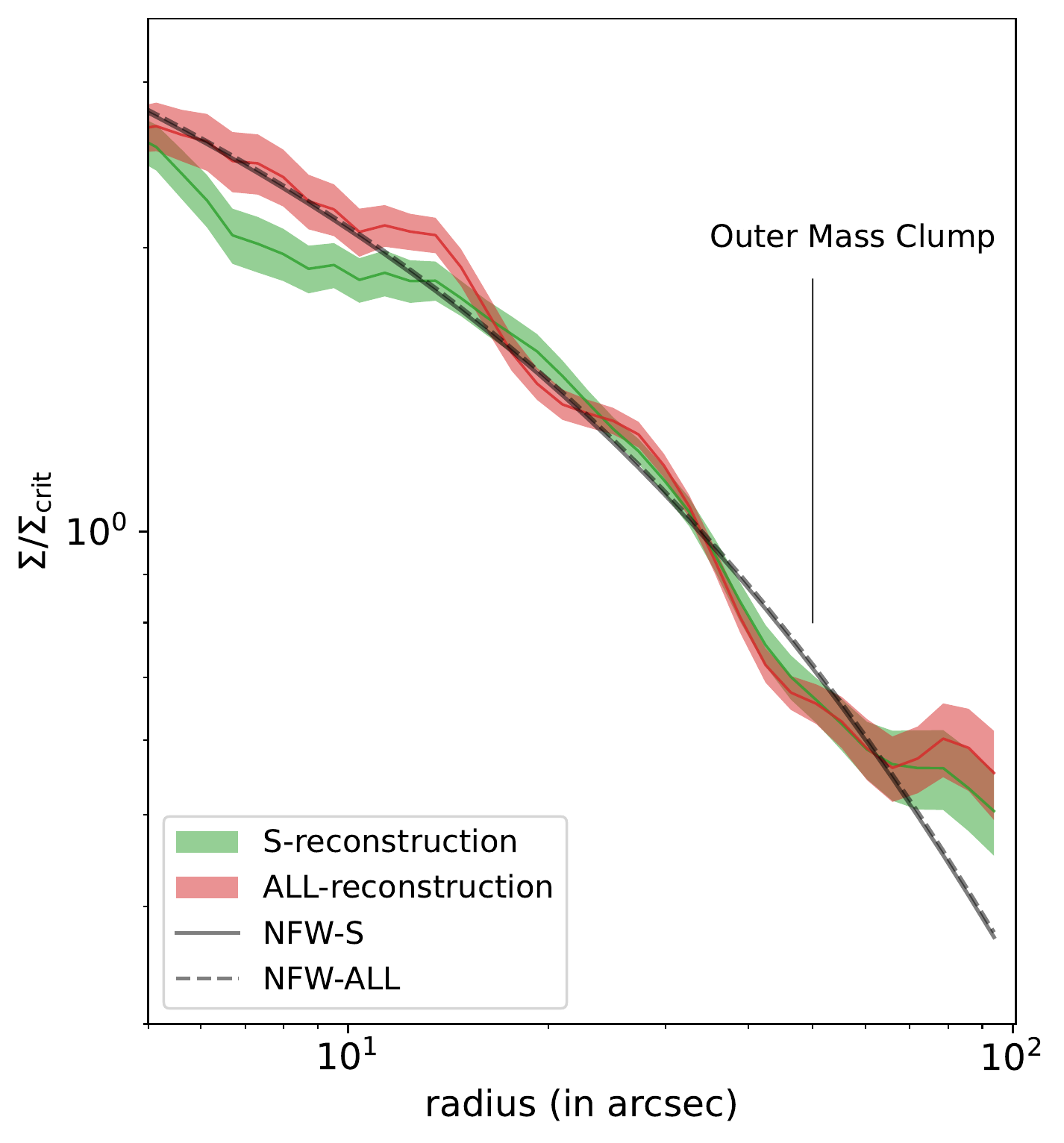}
    \caption{Radial mass profiles of the reconstructions. The best-fit NFW profiles are shown as black dashed line for S-reconstruction, and black dash-dotted line for All-reconstruction. The shaded regions represent the $68\%$ CL uncertainties in the reconstructed mass distributions. The radial profiles are similar for both the reconstructions. The radial location of the outer mass clump in the north-east region is marked in the plot.}
    \label{fig:massprofile_allres}
\end{figure}

Our best fit reconstructions are obtained by averaging 40 independent \textsc{grale} runs. Each run starts with a random seed producing a similar but different map. Constrained by the required computational resources, {the number of independent \textsc{grale} runs} (i.e., 40), is consistent with our previous works \citep{Ghosh2021,Ghosh2020}. In this work, we are using a reconstruction area of $0.62$ Mpc $\times$ $0.62$ Mpc and the smallest resolution cell (projected Plummer sphere) has a radius of $\sim 2$ kpc.

The reconstructed mass distributions for the S- and ALL-reconstructions are shown in Figure~\ref{fig:massmaps}, overlaid over the ACS/F814W mosaic. The surface mass density values in these maps are scaled by $\Sigma_{\rm{crit,0}} =5.47$ kg/m$^2$. The corresponding reconstructed critical lines and caustics for each of the reconstructions are shown in Figure~\ref{fig:magmaps}. In both Figure~\ref{fig:massmaps} and \ref{fig:magmaps} the multiple image systems used to produce the corresponding reconstructions are shown as red stars for systems with spectroscopic redshifts, and red triangles for systems with photometric redshifts. 

The reconstructed mass distribution shows the presence of a prominent mass clump about 50 arcsec ($\sim$155 kpc) to the north-east of the central region. This mass clump is in agreement with all previous reconstructions as well as the previous \textsc{grale} reconstruction obtained by \citet{Mohammed2014}, with a much smaller number of images.

Since the overall mass distribution is approximately circularly symmetric, we compute the circularly averaged density profiles for each reconstruction, centered on the Brightest Cluster Galaxy (BCG). These are shown in Figure~\ref{fig:massprofile_allres}. Both have similar radial profiles. 
Fitting NFW profiles yielded concentration parameter values of $c \sim 8.7$ and scale radii of $R_{\rm S}\sim 100$ arcsec for both reconstructions. These values are consistent with other studies \citep[see table 2 of][]{Coe2010}. {The poor quality of the NFW fit, the relatively short radial range,} $\sim 1.3$ decades in radius, and the presence of significant mass clumps do not allow to differentiate between various theoretical models of dark matter profiles, as was done in \cite{ber13} and \cite{Umetsu16}.

The central $32\times32$ arcsec region shows signatures of possible cases of deviations from mass-following-light. These are examined in detail in Section~\ref{sec:sedresults}.

The regions outside the area covered by lensed multiple images are not well constrained, so the details of that mass distribution are not accurate. However, it is safe to say that the presence of \textsc{grale} reconstructed mass in those regions does mean that there is external mass, but its amount and distance from the cluster center are not certain and are degenerate with each other. \Review{The \textsc{grale} reconstruction of Abell 370 shows similar external masses outside the region covered by the images \citep{Ghosh2021}. Using synthetic clusters, we showed that presence of external masses outside the multiple-image region generates mass clumps in the less-constrained reconstructed region, i.e. the one without multiple images. These masses lead to the extension of the critical curves in the less-constrained part of the reconstructed region.} Our reconstructions of Abell 1689 also show an extension of the mass to the south of the cluster beyond $200$ kpc. \Review{Small mass extensions to the south appear to be present in the reconstructions presented in \citet{Limousin2007} (their fig.~5) and \citet{Cha2022} (their fig.~14).}

\section{Comparison of Lens Mass and Stellar Mass}
\label{sec:SED}
\subsection{SED Fitting with FAST++}
\label{sec:fast}

\begin{figure*}
     \centering
          \begin{subfigure}[b]{0.48\textwidth}
         \centering
         \includegraphics[width=\textwidth]{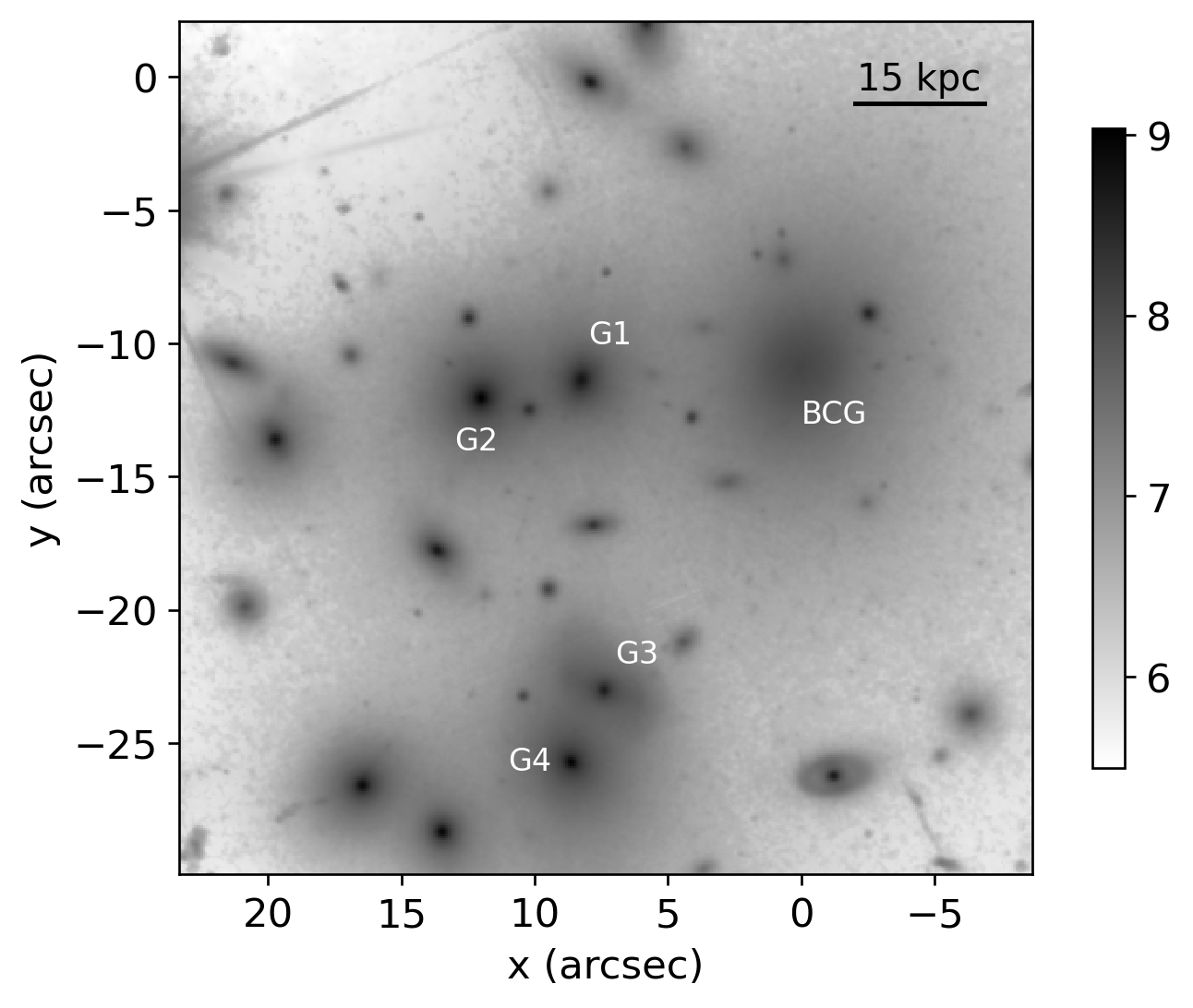}
         \caption{Luminosity distribution in Johnson-B filter}
         \label{fig:johnsonBmap}
     \end{subfigure}
     \quad
     \begin{subfigure}[b]{0.49\textwidth}
         \centering
         \includegraphics[width=\textwidth]{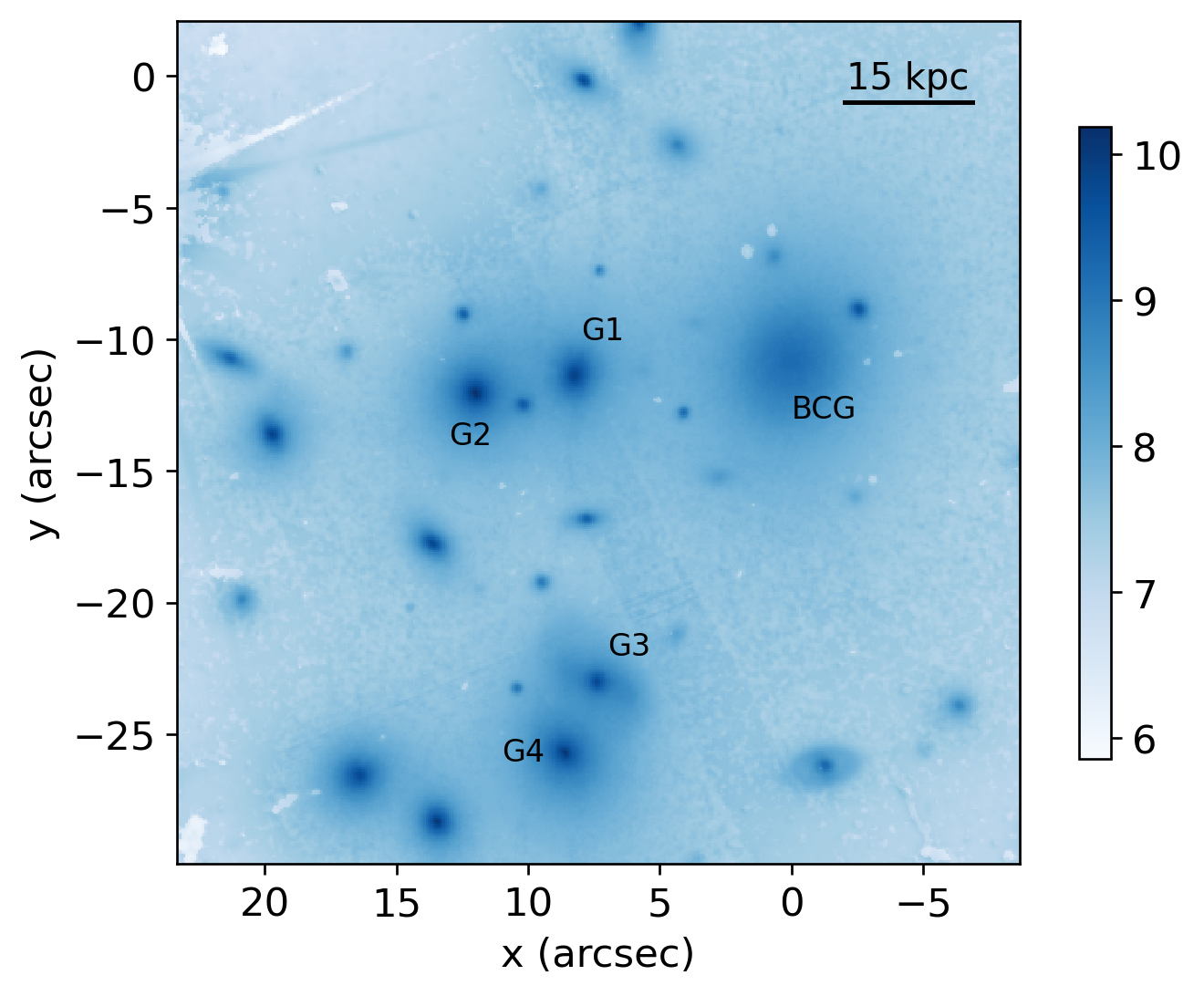}
         \caption{Stellar Mass Distribution}
         \label{fig:stellarmap}
     \end{subfigure}

        \caption{(a) The luminosity distribution in the central $\sim$100 kpc region in the Johnson B-filter in terms of solar luminosities in log scale. \Review{The fluxes in five different HST photometric bands are used as input to \textsc{fast++}.} (b) The stellar mass distribution in the central $\sim$100 kpc region as retrieved by \textsc{fast++} in terms of solar masses in log scale. The BCG and other bright galaxies in this region are marked in black.}
        
\end{figure*}

There are 4 prominent ellipticals in the central $\sim\!100$ kpc region of A1689: BCG, G1, G2, and G4 (see Figure~\ref{fig:stellarmap}). To compute their stellar mass, and stellar mass associated with smaller galaxies in that region we used \textsc{fast++}\footnote{\textsc{fast++} is publicly available on GitHub: \url{https://github.com/cschreib/fastpp}.}, a faster and more efficient version of the original IDL-based code \textsc{fast} \citep{Kriek2009}, that fits Spectral Energy Distribution (SED) templates to broadband photometry and/or spectra. Depending on the input parameters, \textsc{fast++} can compute the best-fit redshift, age, dust content, star formation timescale, metallicity, stellar mass, star formation rate (SFR), and their confidence intervals. To find the best-fit parameters it uses $\chi^2$-minimisation. The confidence levels are calibrated using Monte Carlo simulations.

A  review on various SED fitting methodologies can be found in \citet{Walcher2011} and \citet{Conroy2013}.

We used fluxes in five photometric bands (F336W,  F475W, F625W, F775W and F814W) spanning the observed near-UV and optical. 

The PSF matched and background subtracted flux mosaics in those bands were generated by \citet{Alavi2014} along with the associated flux RMS values. 
However, the flux RMS values only include the instrumental and background errors but exclude the random Poisson noise associated with the flux from the galaxies.
We calculated this Poisson noise as,
\begin{equation}
    \sigma_{\rm p}=\sqrt{{\rm R}\times {\rm t}}=\sqrt{{\rm N}/{\rm G}}
\end{equation}
where R is the electron count rate per second, t is the observation time in seconds, G is the inverse gain, and N is the number of detected events (photons). The calculated Poisson noise values were added in quadrature to the existing RMS maps.

For the input parameters to \textsc{fast++}, we used the \citet{Bruzal2003} stellar population synthesis model library with a \citet{Chabrier2003} IMF and solar metallicity. 
We assume exponentially declining star formation histories, appropriate for ellipticals, with a minimum e-folding time of ${\rm log}_{10}(\tau/{\rm yr}) = 7$, a minimum age of $100$ Myr, $0 < {\rm A_V} < 1$ mag and the \citet{Calzetti2000} dust attenuation law. We fixed the input redshift to the cluster redshift $z_l=0.1832$.

\subsection{Results}
\label{sec:sedresults}

\begin{figure*}
	\includegraphics[width=\textwidth]{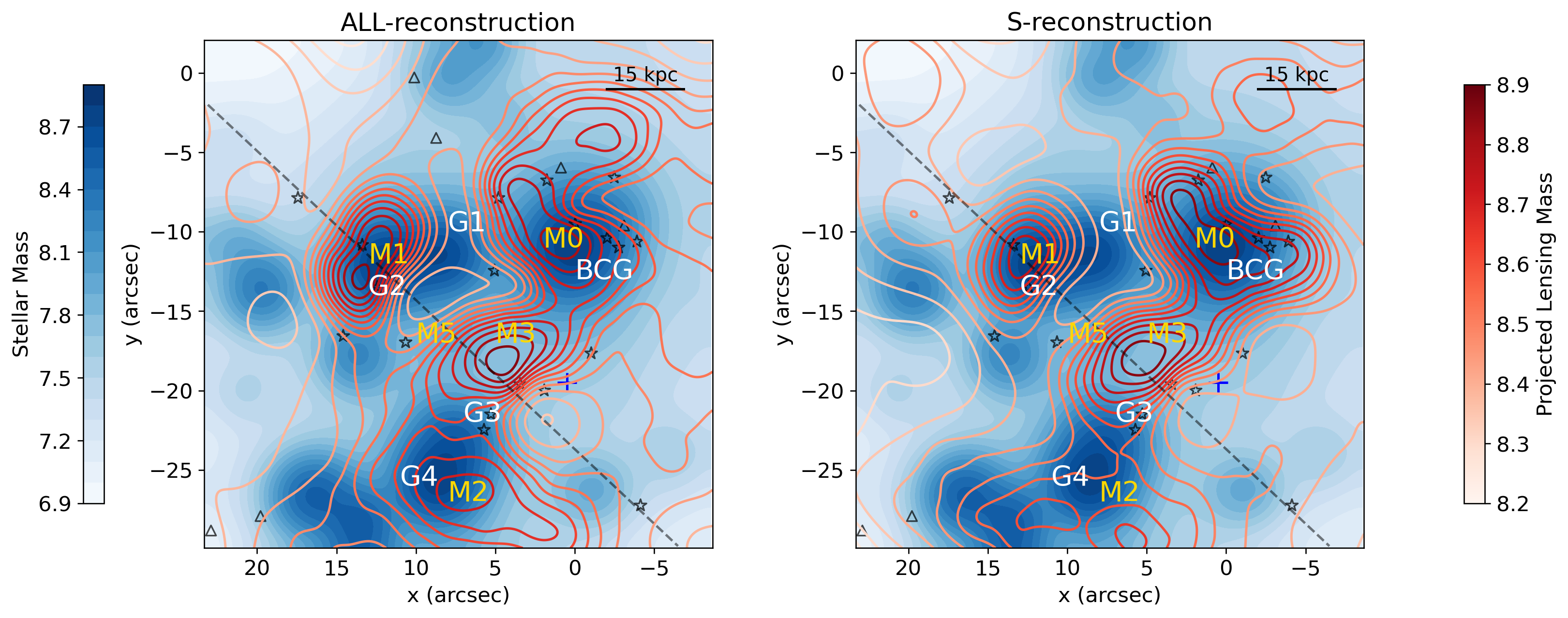}
    \caption{ Reconstructed Projected lensing mass distributions (red contours) from the  ALL-reconstruction (left panel) and S-reconstruction (right panel) overlaid on the stellar mass distribution (blue colour scale) retrieved by \textsc{fast++} in the central $\sim$100 kpc region. Both the reconstructed lensing mass and the stellar mass are in terms of solar masses in log scale per pixel and the pixel size is 80 mas. Yellow markers show the features in the reconstructed mass distribution and the white markers show the BCG and other bright galaxies in the stellar mass distribution as explained in Section~\ref{sec:sedresults}. The stellar mass distribution is smoothed with a Gaussian filter with FWHM of 1.3 arcsec to match the resolution of \textsc{grale}. Blue cross shows the location of the dark matter halo center of \citet{Limousin2007} reconstruction. The cross-section along the dashed black line is shown in Figure~\ref{fig:masscrossprofile_hires}. Black stars and triangles show the position of the multiple images with spectroscopic and photometric redshifts, respectively.}
    \label{fig:mm_hires}
\end{figure*}

\begin{figure*}
	\includegraphics[width=\textwidth]{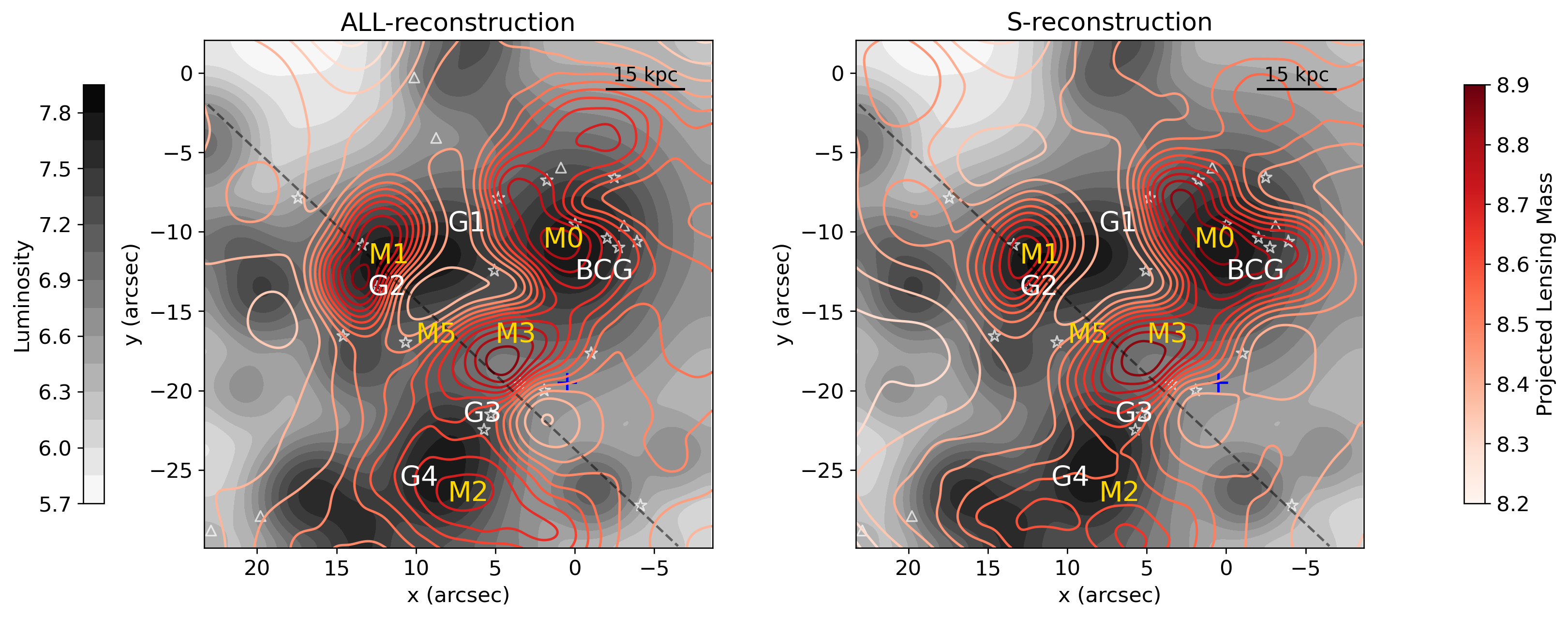}
    \caption{Reconstructed projected  lensing mass distributions (red contours) from the  ALL-reconstruction (left panel) and S-reconstruction (right panel) overlaid on the luminosity distribution (gray colour scale) in Johnson B-filter in the central $\sim$100 kpc region. The reconstructed lensing mass and the luminosity are in terms of solar masses and solar luminosities in log scale per pixel, respectively, and the pixel size is 80 mas. The yellow markers show the features in the reconstructed mass distribution and the white markers show the BCG and other bright galaxies in the luminosity distribution distribution as explained in Section~\ref{sec:sedresults}. The luminosity distribution is smoothed with a Gaussian filter with FWHM of 1.3 arcsec to match the resolution of \textsc{grale}. Blue cross shows the location of the dark matter halo center of \citet{Limousin2007} reconstruction. The cross-section along the dashed black line is shown in Figure~\ref{fig:masscrossprofile_hires}. White stars and triangles show the position of the multiple images with spectroscopic and photometric redshifts, respectively.}
    \label{fig:ml_hires}
\end{figure*}

\begin{figure}
	\includegraphics[width=\columnwidth]{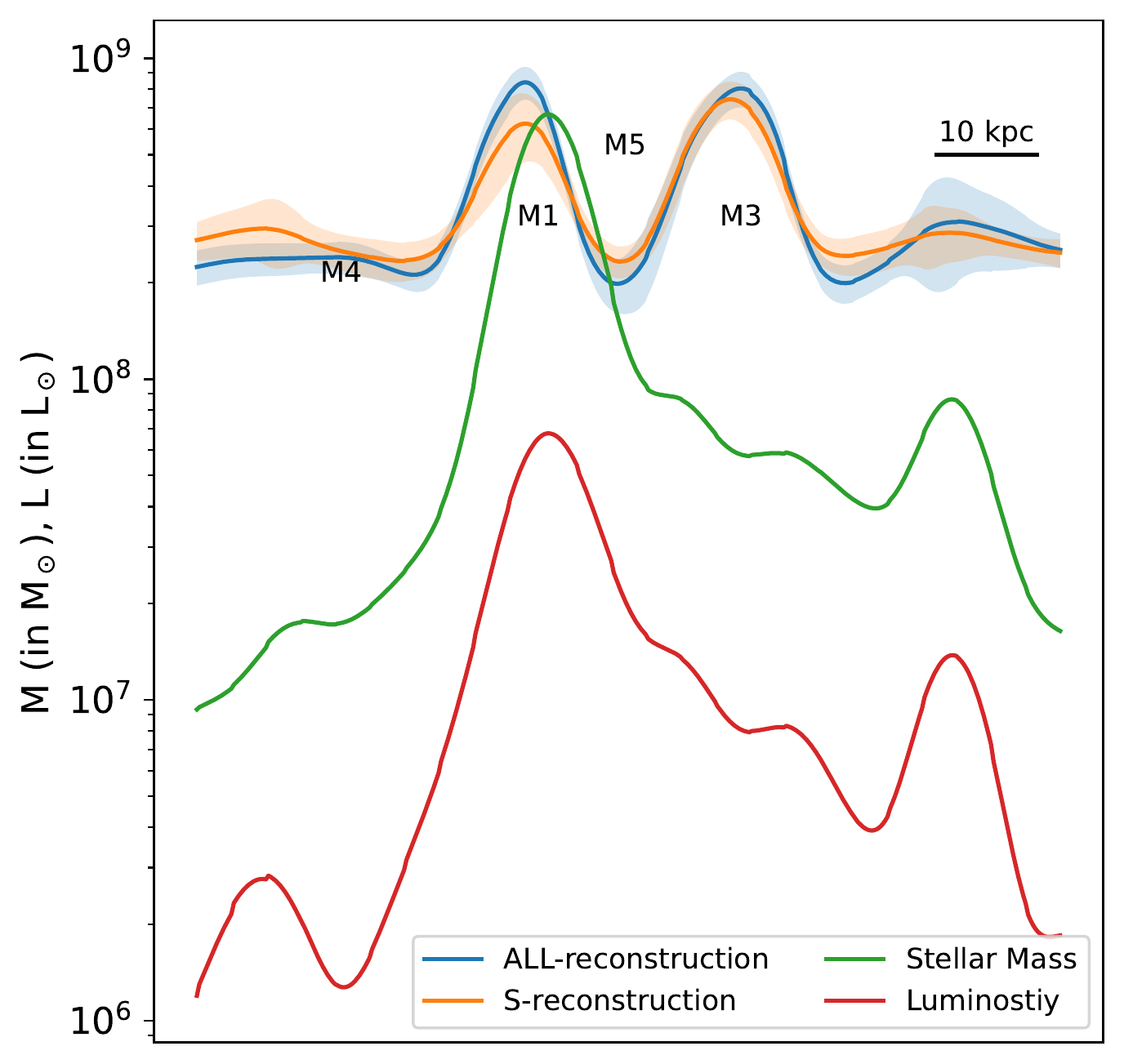}
    \caption{Cross-sectional profiles of the reconstructed projected  lensing mass for both reconstructions, the Gaussian-smoothed stellar mass and the Gaussian-smoothed luminosity along black dashed lines shown in Figure~\ref{fig:mm_hires}. Locations of the reconstructed mass features are marked in black. \Review{The shaded regions around the reconstructed mass profiles show the uncertainties in the reconstructed mass values at 68\% CL as calculated from the 40 \textsc{grale} runs. The M3 peak is highly significant given the uncertainties as shown in the plots.}}
    \label{fig:masscrossprofile_hires}
\end{figure} 

\begin{figure}
	\includegraphics[width=\columnwidth]{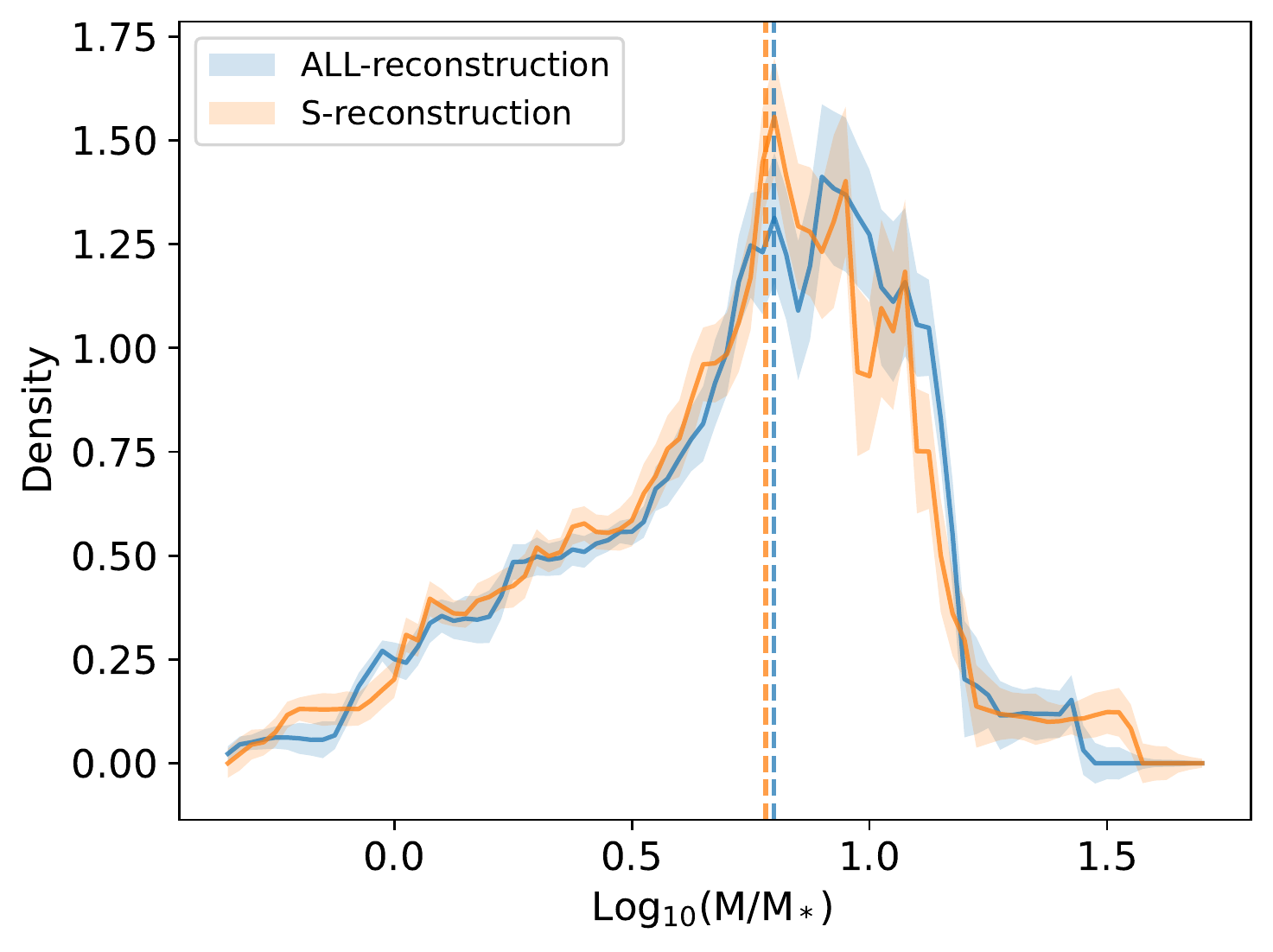}
    \caption{Histograms of the ratio of the total reconstructed lensing mass to Gaussian-smoothed stellar mass (M/M$_*$, in solar units), for both reconstructions in the central $\sim$100 kpc region.  Vertical dashed lines show median values of the ratios. \Review{The shaded regions show the uncertainties in the histogram at 68\% CL as calculated from the 40 \textsc{grale} runs.}}
    \label{fig:mmhist_hires}
\end{figure}

\begin{figure}
	\includegraphics[width=\columnwidth]{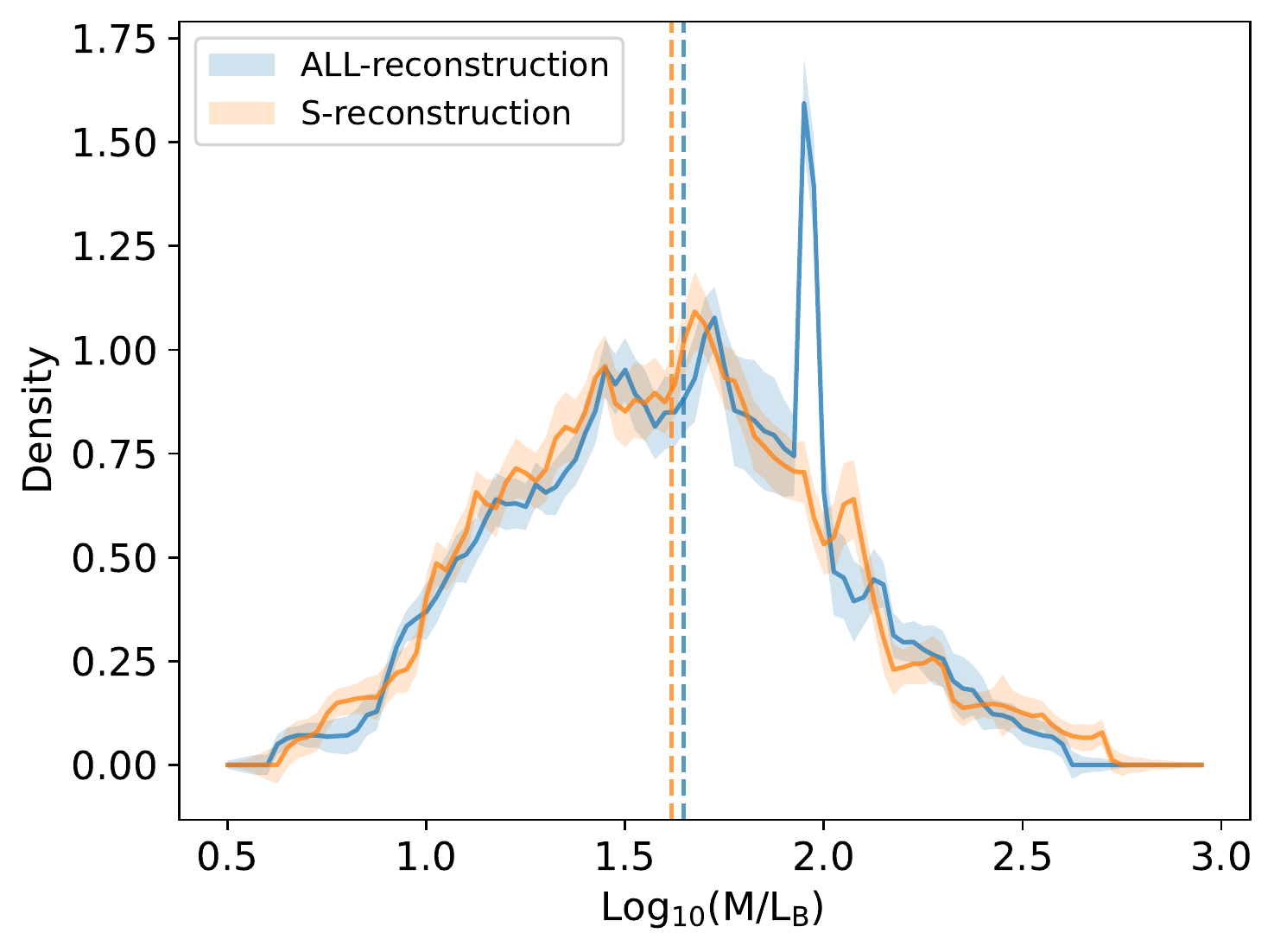}
    \caption{Histograms of the ratio of the total reconstructed lensing mass to Gaussian-smoothed luminosity (M/L$_{\rm B}$, in solar units), for both reconstructions in the central $\sim$100 kpc region. Vertical dashed lines show median values of the ratios. \Review{The shaded regions show the uncertainties in the histogram at 68\% CL as calculated from the 40 \textsc{grale} runs.}}
    \label{fig:mlhist_hires}
\end{figure}

The central $\sim\!100$ kpc region is the most well constrained area in our reconstructions. It contains $\sim\!\!25$ multiple images. The luminosity distribution of that region in Johnson B-filter is shown in Figure~\ref{fig:johnsonBmap}.
Figure~\ref{fig:stellarmap} shows the stellar mass distribution obtained for this region of the cluster in a pixel-by-pixel manner using \textsc{fast++}.\footnote{We did not mask any high-$z$ pixels containing the strongly lensed images, which look like small faded spots in Figure~\ref{fig:stellarmap}. Since they are small and faint, and we are using Gaussian smoothing on the stellar mass distribution (as will be explained later in this section) their contributions are negligible.} The BCG and nearby major bright galaxies (G1-G4) in this region are marked in these figures for easier identification. All these galaxies are ellipticals in nature, with the only exception of G3, which is a spiral.

Since the angular size of the smallest Plummer in our \textsc{grale} reconstructions is about $\sim\!0.66$ arcsec (2 kpc), i.e., larger than the angular resolution of the stellar mass and light maps, in order to do a fair comparison between the two we needed to smooth the stellar mass and light distribution. We used a Gaussian filter on the stellar mass and luminosity maps with an FWHM of $1.3$ arcsec.  

The comparison of the stellar mass distribution to the reconstructed lensing mass distribution is shown in Figure~\ref{fig:mm_hires}, and the comparison of the luminosity distribution to the reconstructed lensing mass distribution is shown in Figure~\ref{fig:ml_hires}. {Figure~\ref{fig:masscrossprofile_hires} shows a cross-sectional profile of the total and stellar mass, as well as stellar light, along the dashed line in Figures~\ref{fig:mm_hires} and \ref{fig:ml_hires}.} The fact that this region is well constrained by lensed images provides confidence to the reality of the mass structures. In this region we found three lensing mass peaks, M0, M1, M2, all of which are accompanied by stellar counterparts, BCG, G1+G2, G4, respectively. Galaxies G1 and G2 are 3.7 arcsec apart, making it hard for \textsc{grale} to resolve. 

The mass clump associated with the BCG is diffuse, similar to its light distribution. There is a slight offset between the BCG and the nearest reconstructed mass peak, M0. \Review{The offset is not statistically significant for these high resolution reconstructions, but are somewhat significant for lower resolution ones; see Appendix~\ref{sec:lores}. For both sets of reconstructions, our offsets are not at all unusual given those seen between X-ray and velocity centers of lower redshift galaxy clusters \citep{Lauer2014}.}


Our lensing reconstructions show a mass peak M3, which is not accompanied by any light peaks. Between M1 and M3, there is a mass dip, M5, which is $\sim\!15$ kpc wide. 

\Review{Figure~\ref{fig:masscrossprofile_hires} shows the cross-sectional profiles of both the reconstructed peak M3 along with the uncertainties in the reconstructed mass at 68\% CL as calculated from the 40 different \textsc{grale} runs. As seen from the plot the light-unaccompanied mass peak M3 has very low uncertainty values for the reconstructed mass, which is an indicator that all 40 \textsc{grale} runs have reproduced this peak consistently.} 
{We examine M3 in more detail in Section~\ref{sec:M3}.}
Mass peak M1 is very near the corresponding galaxy G2 and its lensing recovered mass is similar to the stellar mass within $\sim 5$kpc radius where stellar mass is expected to dominate.

The histogram of the ratio of the total mass to stellar mass is shown in Figure~\ref{fig:mmhist_hires}. Median M/M$_*$ values are 6.3 and 6.05 for ALL- and S-reconstructions, respectively.  M/M$_*$ is greater than 1 for most of the region, which is reassuring because the lensing mass should be greater than the stellar mass in most of the region. We stress that \textsc{grale} does not use cluster galaxies as input, so this is an unbiased measure of the mass distribution in the very center of the cluster.

The histogram of the ratio of the total mass to light (in solar units) is shown in Figure~\ref{fig:mlhist_hires}. Median M/L$_{\rm B}$ values are 41.1 and 41.3 for ALL- and S-reconstructions, respectively. M/L$_{\rm B}>10$ for most of the region. 

The stellar mass-to-light ratio is $\sim\!10$ for most of the central  100 kpc region (see green and red curves in Figure~\ref{fig:masscrossprofile_hires}), which is consistent with, but on the upper end of values in the literature.  \cite{and19} analyzed Abell 611, and found its stellar M/L$_{\rm B}\sim 3-10$, which they consider high. A follow-up study \citep{and22} analyzed eight strong lensing clusters. One cluster, MACS2129 had M/L$_{\rm B}\sim 10$, while the other seven (including A611) had M/L$_{\rm B}\sim 1-4$. For ellipticals outside of galaxy clusters M/L ratios are similar or lower. A sample of SAURON nearby ellipticals has stellar M/L$_{\rm B}$ in the range $1-3$, within their effective radii \citep{cap06}, while a study of 7 ellipticals by \cite{hum06} obtained M/L$_{\rm B}<1$ interior to galaxies' effective radii. 

\subsection{A closer examination of recovered mass peak M3}\label{sec:M3}

{Because the light-unaccompanied mass clump M3 needs more scrutiny,} we searched for light signatures from dust-obscured galaxies in the {\it Spitzer/IRAC} data of the same region in all four {\it IRAC} infra-red bands, but were unable to find any such structure, while the other galaxies in the region showed up prominently. {\cite{hai10} obtained {\it Herschel/PACS} 100 $\mu$m map of A1689: there is no spectroscopically confirmed cluster member at the location of M3.} This eliminates the possibility of the presence of any dust-obscured galaxies associated with M3.

{Figures~\ref{fig:lprmshist} and \ref{fig:lprmshist_lowres} show that the typical lens plane rms is well above HST astrometric precision, indicating that \grale\ reconstructions are not suffering from overfitting, and so M3 is unlikely to be the result of overfitting. Furthermore, in \cite{Ghosh2020}, using \grale\ reconstructions of mock clusters, we showed that in lens plane regions with image number density comparable to the central region of A1689 the fractional deviation between the true and recovered maps is <30\%, suggesting that M3 is a real feature. }

The location of M3 roughly matches with the location of dark matter halo center of \citet{Limousin2007}, indicated by the blue cross in Figure~\ref{fig:mm_hires}. {However, the radius of the flat density core of their dark matter halo is very large, $\sim 100$kpc, which means that the center could have been many kiloparsecs away.} The reconstruction obtained by \citet{Coe2010} has an elongated structure in that region stretching out from the mass clump associated with the BCG. {These two studies (simply-parametrized \lenstool and free-form \textsc{LensPerfect}) may suggest that our M3 is similar to mass clumps detected by other reconstructions, but the evidence is not very strong.}

\begin{figure*}
    \centering
    \includegraphics[trim={0cm 5cm 0cm 4cm},clip,width=0.49\textwidth]{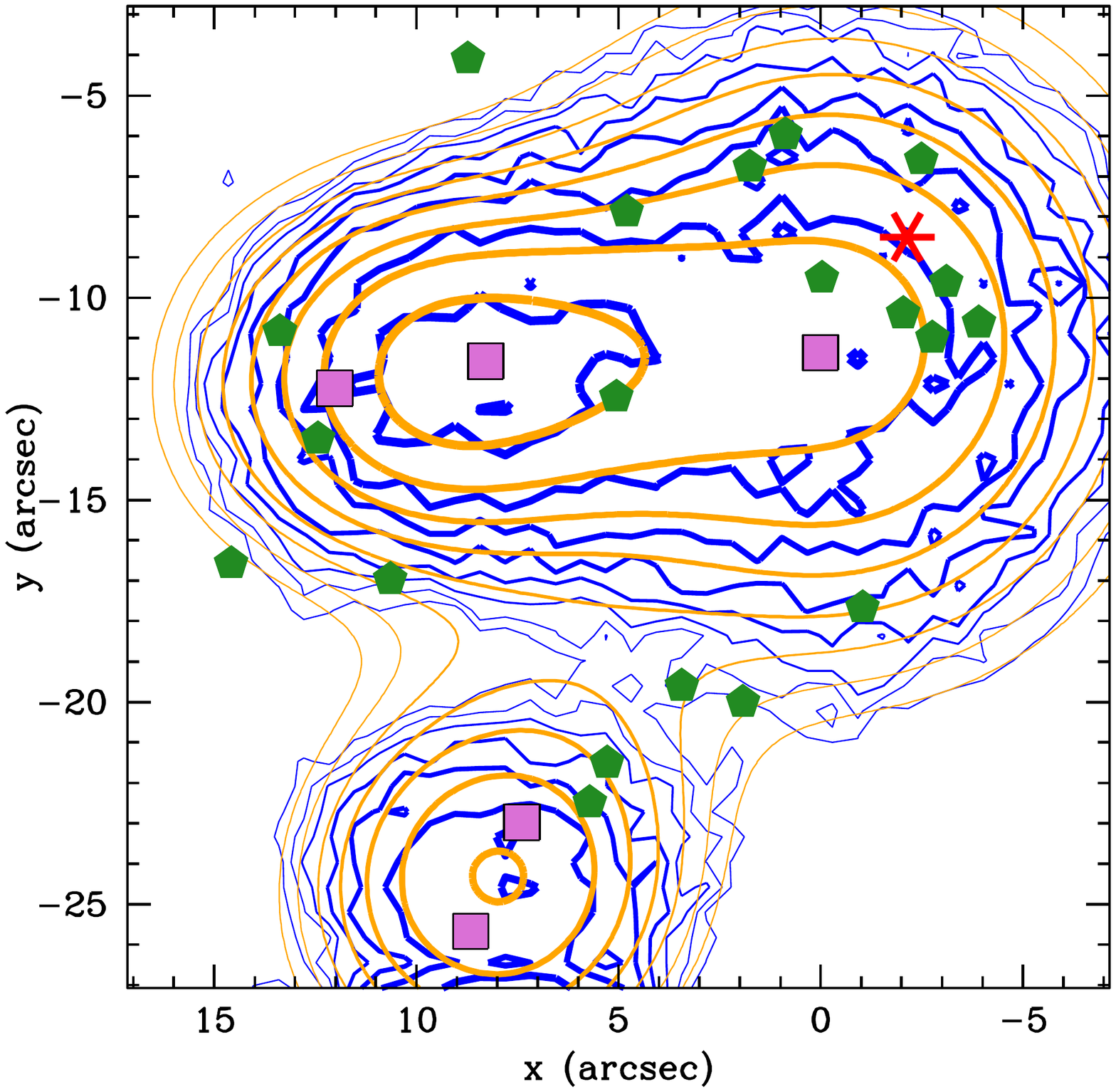}
    \includegraphics[trim={0cm 5cm 0cm 4cm},clip,width=0.49\textwidth]{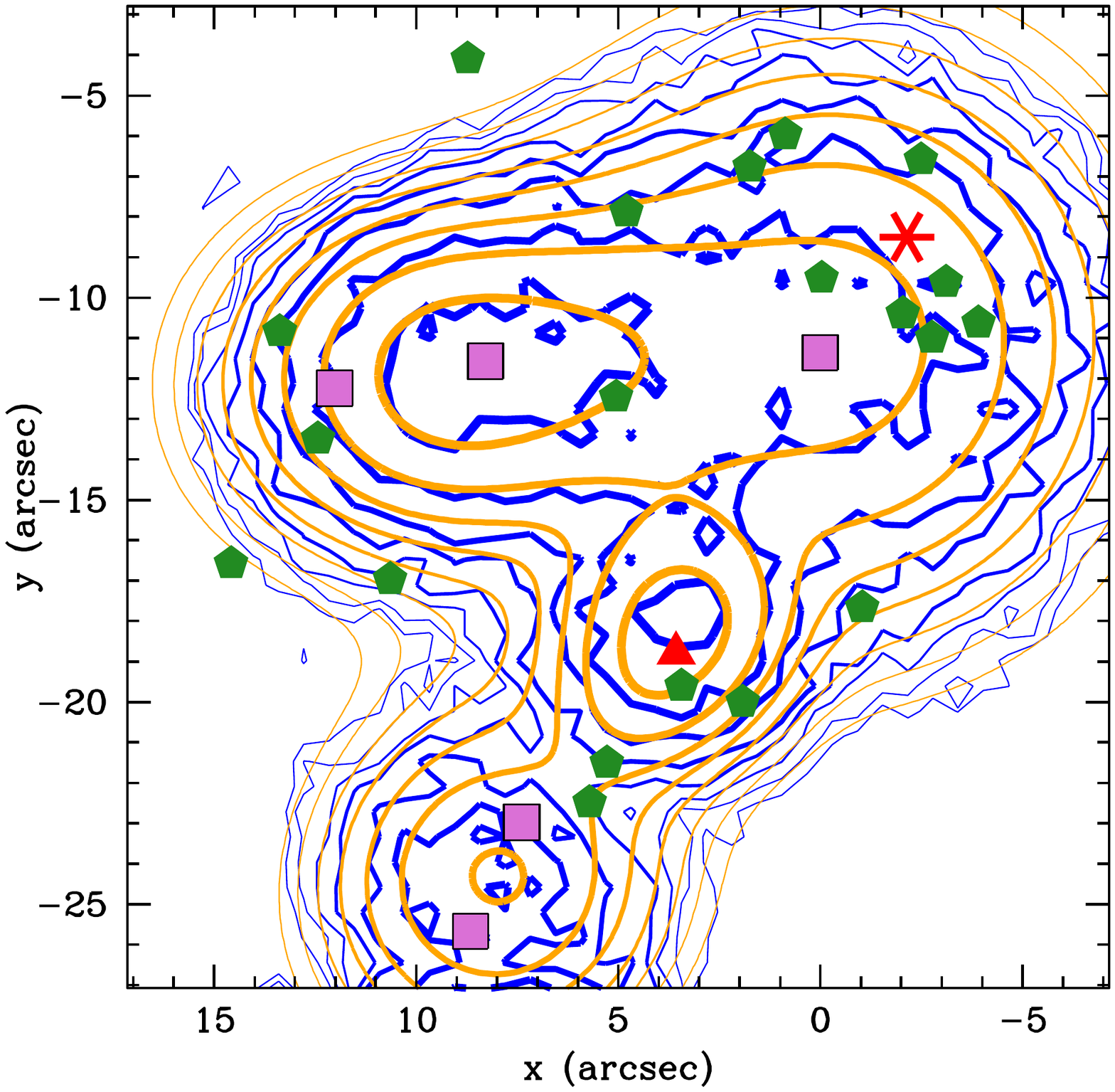}
    \caption{{A toy model of the central region of A1689, which includes the main dark matter halo (centered on the red asterisk), and 5 galaxies (BCG, and G1-G4), represented by pink squares. Observed images are marked with green pentagons. Orange contours are the projected surface density of the lens, while the blue contours show the probability distribution of the number density of maxima images generated by a very large number of sources.   The mass density contours (orange) are spaced by factors of $0.9$, while those of the image number density (blue) by factors of $0.72$, so the images are more concentrated, or clustered than the mass. The two sets of contours track each other closely. {\it Left:} No additional dark matter clump. {\it Right:} This model includes an additional dark matter clump (centered on the red triangle) approximating M3. Comparison of the two panels shows that some mass concentration is needed near the red triangle to produce nearby images.}}
    \label{fig:mock1}
\end{figure*}

\begin{figure*}
    \centering
    \includegraphics[trim={0cm 5cm 0cm 4cm},clip,width=0.49\textwidth]{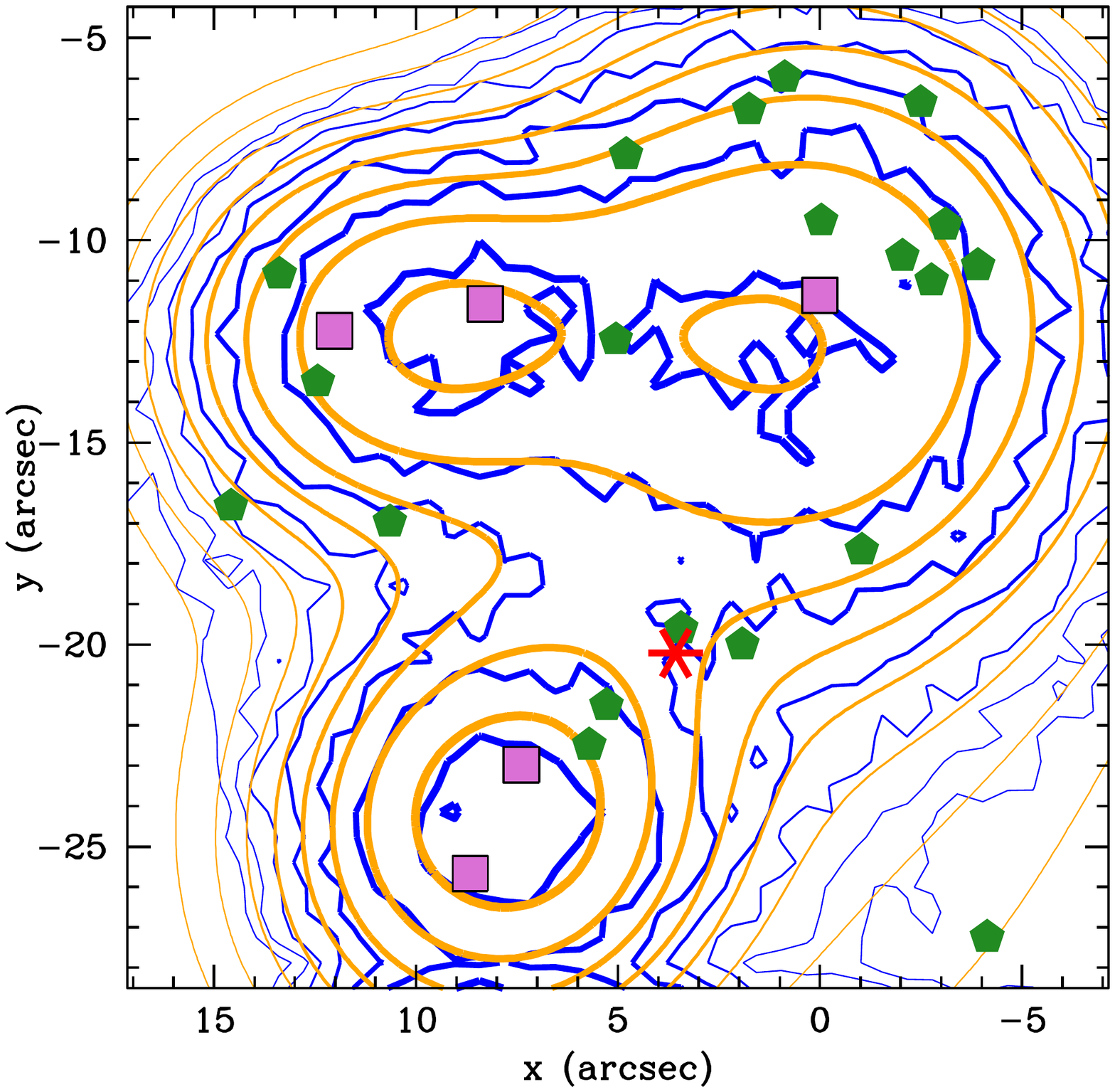}
    \includegraphics[trim={0cm 5cm 0cm 4cm},clip,width=0.49\textwidth]{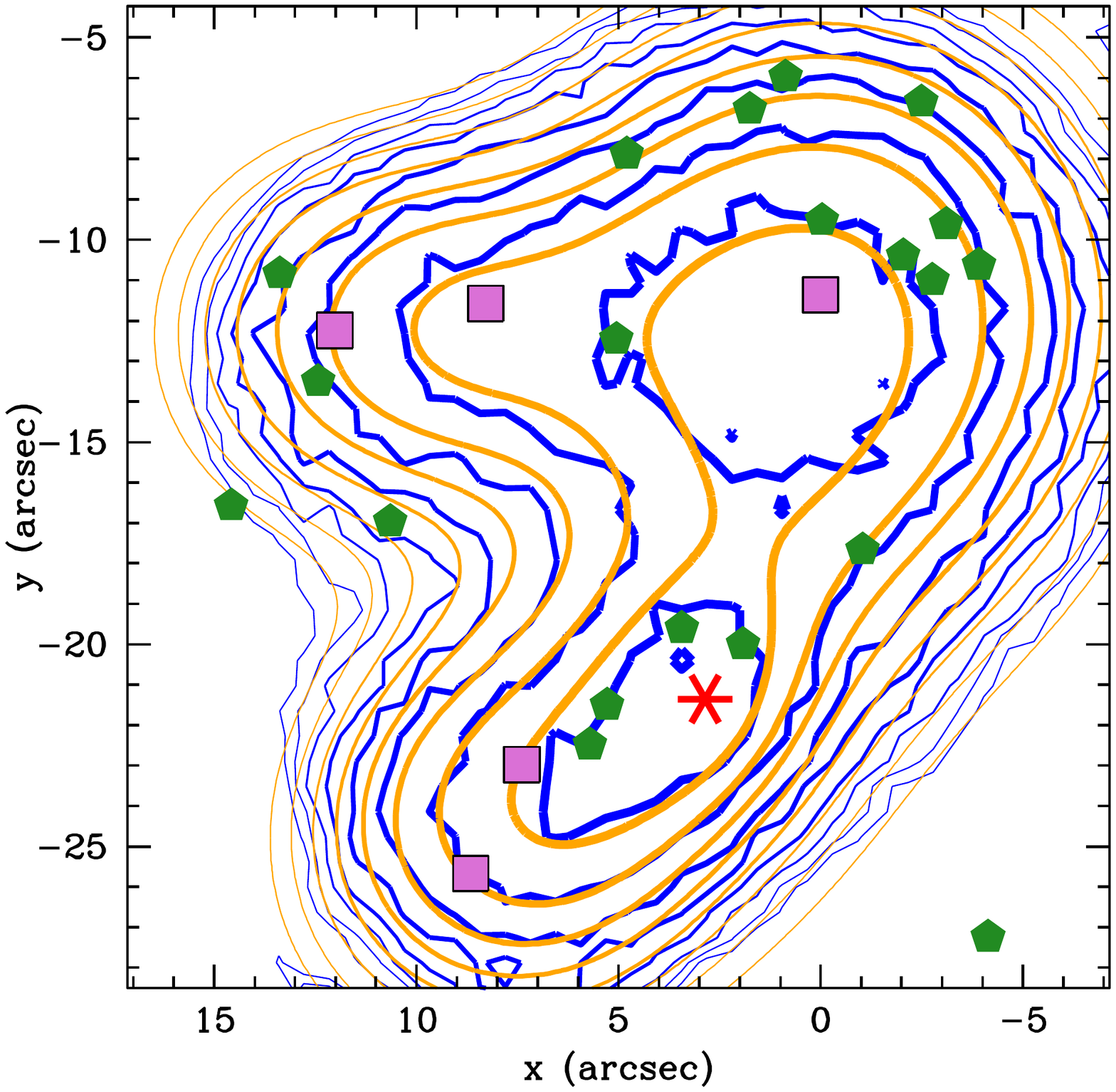}
    \caption{{Similar to Figure~\ref{fig:mock1}, but no additional mass clumps are included in either of the panels. Instead, the center of the main dark matter halo is moved to be near M3, and its center is indicated by the red asterisk. {\it Left:} The dark matter halo has a large density core, $r_c=16''$, or 50 kpc. This scenario is less likely to generate the 2-5 images near $(3'',-20'')$. {\it Right:} The dark matter halo has a small density core, $r_c=4''$, or 12 kpc. In this case the maxima images are generated in the region in question, since there is still a local mass peak at that location.}}
    \label{fig:mock2}
\end{figure*}

{Here we present an argument that could support the reality of M3. It is based on the locations of the 16 images in the central $\sim 24''$ region of A1689. All these images are maxima in the arrival time (Fermat) potential, which is determined by the mass distribution and the source position. Their identity as maxima is indicated by the \grale\ reconstructions. Other models, for example \cite{Halkola2006}  and \cite{Diego2015}, also identify these as maxima. Even without a mass model one can say with reasonable confidence that these are maxima. Maxima images are characterized by two negative eigenvalues of the magnification matrix, $(1-\kappa-\gamma)$ and $(1-\kappa+\gamma)$. In the central region of A1689 $\kappa$ is large and $\gamma$ is relatively small due to the flat density core of the cluster, so both eigenvalues are likely negative.}

{It is well known that maxima tend to form close to the local density peaks. We demonstrate that there is an even closer association between the two: in the absence of strong external shear the probability distribution of the surface number density of maxima follows very closely that of the surface mass density of the lens.}

{The toy model we use in this subsection is not based on any specific model of A1689. Instead, we work with a rough parametric mock of the central $\sim 24''$ region, which includes the main dark matter halo, represented by an elliptical {\tt alphapot} potential \citep{Keeton2001} with a core radius of $r_c=16''$, and 5 galaxies, the BCG, and G1-G4, represented by circular Einasto profiles \citep{Dhar2022} of index 3, scale radius $r_S$, and relative normalizations that approximate the relative brightnesses of these galaxies. The mocks are not meant to give an accurate map of A1689, but only to capture its main features.  We tried several sets of these parameters; one example is shown in Figure~\ref{fig:mock1}. The red asterisk is the center of the main dark matter halo. Pink squares and green pentagons represent the 5 galaxies and observed images, respectively. Orange contours show the projected surface mass density, while the blue contours the number density of maxima images. To generate the blue contours we populated the source plane with few $\times 10^5$ sources (all at the same redshift), forward lensed them and kept only the maxima. Such a high number of sources allowed us to plot the surface number density of these images, i.e., their probability distribution in the lens plane.}

{One can see that the contours of the projected mass density and those of the number density of maxima track each other closely, implying that the observed number density of maxima is an indicator of where the local mass peaks are. The left panel does not include any extra mass clumps, while the right panel does (red triangle), roughly at the location of M3 as recovered by \grale. We did not attempt to refine our mocks to accurately reproduce the observed image number density at all locations in the lens plane, because we are mostly interested in the 2-5 images around $(3'',-20'')$, which do not immediately correspond to any galaxy. Comparison of the two panels of Figure~\ref{fig:mock1} shows that an extra mass clump needs to be present near the location of the red triangle to account for these images. No galaxy is present near that location.}  

{One may ask if centering the main dark matter halo near $(3'',-20'')$ could generate the maxima images in that vicinity. The left panel of Figure~\ref{fig:mock2} shows such a configuration, with the red asterisk marking the location of the large dark matter halo that has a large density core. As before the mass density and image number density track each other well, and there is no excess of maxima outside of the regions directly associated with visible galaxies.} 

{An alternative explanation for M3 is that the main cluster halo has a small core, $r_c=4''$, or $\sim 12$~kpc, and is centered near $(3'',-20'')$. That would also result in a mass density peak at that location and generate maxima images nearby, as shown in the right panel of Figure~\ref{fig:mock2}. However, such a small core appears to be inconsistent with A1689 and some other clusters \citep{lim22}.}

{The above exercise offers support for the hypothesis that there is extra mass at the location of M3, which is not associated with any light. However, we caution that more data and reconstruction work by different methods are needed before one can state with confidence that M3 is real. Because our analysis depends critically on the central maxima images, it is important to confirm them observationally, especially since maxima are demagnified and appear faint.}

Assuming that M3 is real, we can estimate some of its properties. From Figures~\ref{fig:mm_hires} and \ref{fig:masscrossprofile_hires}, M3's projected density is at least $\Delta\kappa\approx 0.4$ above the nearby background projected density. ($\kappa$ is surface mass density in units of critical surface mass density for lensing, $\Sigma_{\rm{crit,0}}$.) Its radial extent on the sky is $r\sim 5\,$kpc. Assuming a similar extent along the line of sight, M3's mass is $\sim 8.3\times 10^{10} M_\odot$, somewhat smaller than that of an $L_*$ galaxy. Its 3D mass density is $n\sim 1.1\times 10^{-23}\,$g/cm$^3$. 

If M3 is a localised mass clump instead of an extended mass elongated along the line of sight, there are two possibilities for the dynamical state of M3 (an intermediate state is also possible). If M3 is gravitationally self-bound, then $GM/r\sim v^2$ would imply that a typical velocity of the particles making it up is $v\sim 270\,\,$km/s. If M3 is not bound, then typical particle velocities within it are similar to those in the center of the cluster, $\sim 1500\,\,$km/s, \citep{Lemze2009} and the structure would disperse in a time it would take a particle travelling at typical velocity to cross it, $\sim 6 \times10^6\,$years. Since this is a very short time scale, we conclude that either we are observing the cluster at a very special time, or M3 is gravitationally self-bound.

Assuming the latter, and if dark matter is self-interacting, with cross-section $\sigma$ cm$^2$/g, the time scale for the structure to disperse due to particles scattering off each other is, 
$\tau\sim (n\,\sigma\,v)^{-1}$ \citep{mir02,pet13}. This can be rewritten as, 
$\tau=1.27\times 10^7 (r/{\rm kpc})^{1/2} (\sigma/[{\rm cm}^2/{\rm g}])^{-1} (\Delta\kappa)^{-3/2}\,$ years. Assuming $\sigma=1$ cm$^2$/g, yields a time scale of $\tau\sim 1.1\times 10^8\,$years. If the cluster formed at $z=0.5$, then that value of $\tau$ represents $\sim3.8\%$ of the time between $z=0.5$ and the redshift we observe it at, $z=0.18$. 
The shortness of $\tau$ suggests that the actual cross-section is smaller, which would allow M3 to persist for a longer time. This limit is similar to some other limits from the literature: based on the Bullet Cluster, \cite{Robertson2017} estimate self-interaction cross-sections $\sigma\lesssim 2$ cm$^2$/g. \cite{Harvey19} find that $\sigma>0.12$ cm$^2$/g are probably ruled out. A recent analysis of eight strong lensing galaxy clusters by \cite{and22} concludes that $\sigma$ is consistent with 0. Though our limit is not as stringent as that, it does suggest that dark mass clumps within clusters can yield constraints on dark matter self-interaction properties.

\section{Discussion and Conclusions}
\label{sec:discussion}

We have obtained reconstructions of Abell 1689 using our free-form algorithm \textsc{grale} with two different sets of multiply imaged systems from the available data. The first set, which contains 107 images, consists of only the secure systems - the S-reconstruction - and the second sets, with 151 images, consists of all available systems excluding a few problematic systems (see Section~\ref{sec:input}). Both of the reconstructions produced similar mass distributions (see Figure~\ref{fig:massmaps}) and their circularly averaged density profile agrees very well with each other as shown in Figure~\ref{fig:massprofile_allres}. {We fit the circularly averaged density profiles with the NFW profile in the range $\sim 15-100\,$kpc, obtaining  concentration parameter values of $c \sim 8.7$,} which are in agreement with prior reconstructions of A1689 (see Section~\ref{sec:reconstructionresults}). \Review{The lensing reconstructed mass profiles} do not indicate the presence of a flat density core down to $\sim 15\,$kpc. The absence of a core $\gtrsim 50\,$kpc is somewhat at odds with the findings of \cite{lim22} who fit parametric cored and non-cored profiles to three other clusters: AS 1063, MACS J0416, and MACS J1206, and found that cored models fit better. However, in MACS J0717 \cite{Limousin2016} found that cored and steep central profiles fit equally well.

In this work we focus on two separate scales of the reconstruction results: mass features inside the central $\sim$100 kpc region, and the mass features outside the central $\sim$100 kpc region. On the latter scale, we found two main mass features 
(see Section~\ref{sec:reconstructionresults}): first, the outer clump in the north east region, about 50 arcsec (155 kpc) from the cluster center which is retrieved by every previous reconstruction of this cluster by different lens modelling groups using different parametric and free-form methods.
Second, we found a southern extension of the overall mass distribution (beyond the image region, $\sim 200$ kpc) of the cluster. \Review{Small mass extensions to the south appear to be present in the reconstructions presented in \citet{Limousin2007} (their fig.~5) and \citet{Cha2022} (their fig.~14), and may suggest the presence of some external mass in that direction.}

For the central $\sim 100\,$kpc, we computed the stellar mass distribution using the SED fitting software \textsc{fast++} and fluxes in photometric bands spanning the near-UV and optical (see Section~\ref{sec:fast}). This region is the most well constrained area within the reconstructions containing $\sim 25$ multiple images. Comparison of the stellar mass with the \textsc{grale} reconstructed lensing mass shows three lensing mass peaks accompanied with bright stellar counterparts (see Figure~\ref{fig:mm_hires}). The lens-recovered mass clump associated with the BCG is diffuse, similar to its light distribution (see Figure~\ref{fig:johnsonBmap}). 

There is a lens-reconstructed mass peak, M3, near the center of this region that does not have a stellar counterpart (i.e., it is dark). Our inspection of \textit{Spitzer} infra-red data found no presence of any dust-obscured galaxies in that region, {and {\it Herschel} does not have any confirmed cluster members at that location.}
{To test the reality of M3 we constructed a toy model that uses local maxima images in the Fermat potential. We show that the contours of the probability distribution of the number density of maxima images and those of the mass density distribution track each other closely. Therefore the number density of observed maxima is a guide to the distribution of the local mass peaks. The importance of the maxima in free form methods was also one of the conclusions in \cite{Massey2018}. In A1689, the need for extra mass near M3 is supported by the distribution of these images in the vicinity of M3, but further data and additional reconstructions are needed to ascertain the existence of this structure. Specifically, it is important to confirm observationally the maxima images that our analysis relies upon. Lens reconstructions by other methods that use these images will also be very helpful.} If confirmed with future analyses, this dark peak can be interpreted in terms of dark matter properties as discussed in Section~\ref{sec:sedresults}. 

Since \textsc{grale} does not use any cluster light distribution as input, it provides a measure of the mass distribution in the central part of the cluster, which is unbiased by the baryonic mass. The median of the ratio of stellar mass to lensing mass ($\mathrm{M/M_*}$) in this region are found to be $\sim 6$ for the S- and ALL- reconstructions (see Figure~\ref{fig:mmhist_hires}). The $\mathrm{M/M_*}>1$ for most of the region is in agreement with the fact that the lensing mass should be greater than the stellar mass.  We found median mass-to-light ratios ($\mathrm{M/L_B}$) in this region to be $\sim 41$ for both reconstructions, respectively (see Figure~\ref{fig:mlhist_hires}), which is consistent with, but on the upper end of, values found in the literature. 

\Review{The lens plane angular resolution in \grale\ can be adjusted by the modeler. In the main paper we have carried out all the reconstructions and analysis using high resolution reconstructions. The lower resolution, which is the standard resolution used in the published \grale\ papers \citep[e.g.,][]{Ghosh2021,Ghosh2020,seb19,williams2019} is presented in Appendix~\ref{sec:lores}. The two sets of results agree very well in terms of reproducing the mass distribution, including the significance of the M3 dark mass peak. }

The goal of lensing reconstructions is to produce maps with properly quantified accuracy and precision. Comparison between reconstructions that use different lens inversion methods is probably the best way to accomplish that. In \cite{Ghosh2021} we carried out such a comparison for the merging cluster Abell 370, based on the data from Hubble Frontier Fields (HFF) and Beyond Ultra-deep Frontier Fields and Legacy Observations (BUFFALO) strong lensing data sets. For this work on A1689 we conclude that on spatial scales of about $100\,$kpc and above, most, and probably all mass reconstructions agree. The discrepancies between reconstructions start on scales below $\sim 100\,$kpc. This is consistent with our conclusions from Abell 370, where \textsc{grale} found an offset of $\sim$35 kpc between the BCG and the associated reconstructed mass peak. A survey of the literature on A370 showed that 11 out of 16 reconstructions had a roughly similar offset. We also found a mass clump of size $\sim\!$ 100 kpc, $\sim\!$ 250 kpc to the east of the A370 cluster center. Out of the same 16 reconstructions, 7 has a roughly similar feature. It is encouraging that reconstructions that use different methodologies often agree on such mass features. When evidence for these becomes more convincing, they can be used to place stringent constraints on the nature of dark matter.

As a future prospect, we plan to incorporate the stellar mass distribution from the SED fitting results as a starting point for the \textsc{grale} based solutions.

\section*{Acknowledgements}

The authors would like to thank Najmeh Emami, Lucy Fortson, Tod Lauer, Marceau Limousin, Jack Sayers and Brian Siana for useful suggestions and discussions. AG acknowledges the School of Physics and Astronomy, University of Minnesota for partially supporting this work through the Allen M. Goldman Fellowship. AG and LLRW acknowledge the support and the computational resources provided by Minnesota Supercomputing Institute (MSI) which were crucial for this work. JL acknowledges the use of the computational resources and services provided by the VSC (Flemish Supercomputer Center), funded by the Research Foundation - Flanders (FWO) and the Flemish Government.


\section*{Data Availability}

The data products associated with this article will be shared on reasonable request to the corresponding authors.




\bibliographystyle{mnras}
\bibliography{ref}



\appendix

\section{Low Resolution Reconstructions}\label{sec:lores}

\begin{figure*}
     \centering
     
     \begin{subfigure}[b]{0.49\textwidth}
         \centering
         \includegraphics[width=\textwidth]{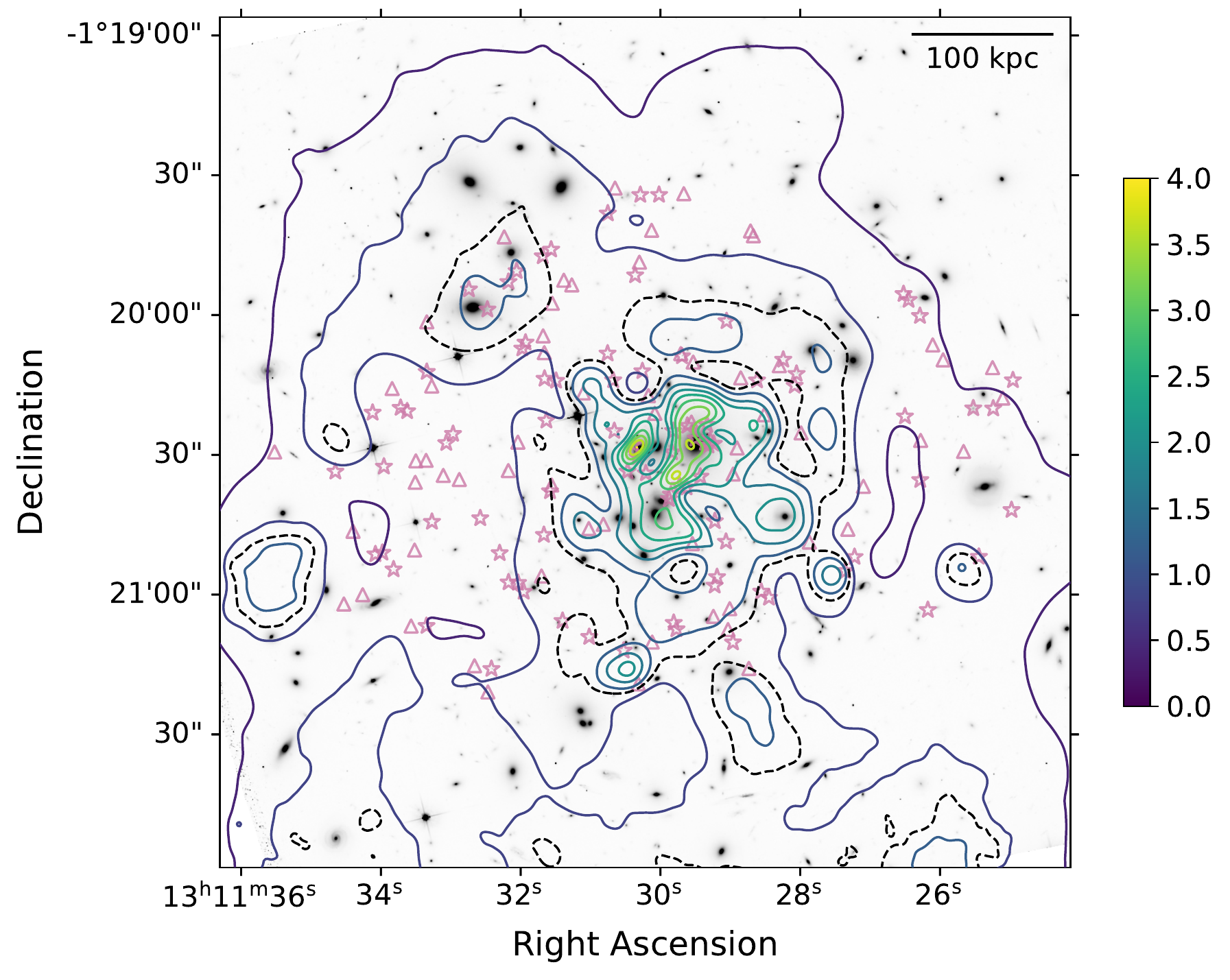}
         \caption{ALL-reconstruction}
     \end{subfigure}
     \quad
     \begin{subfigure}[b]{0.49\textwidth}
         \centering
         \includegraphics[width=\textwidth]{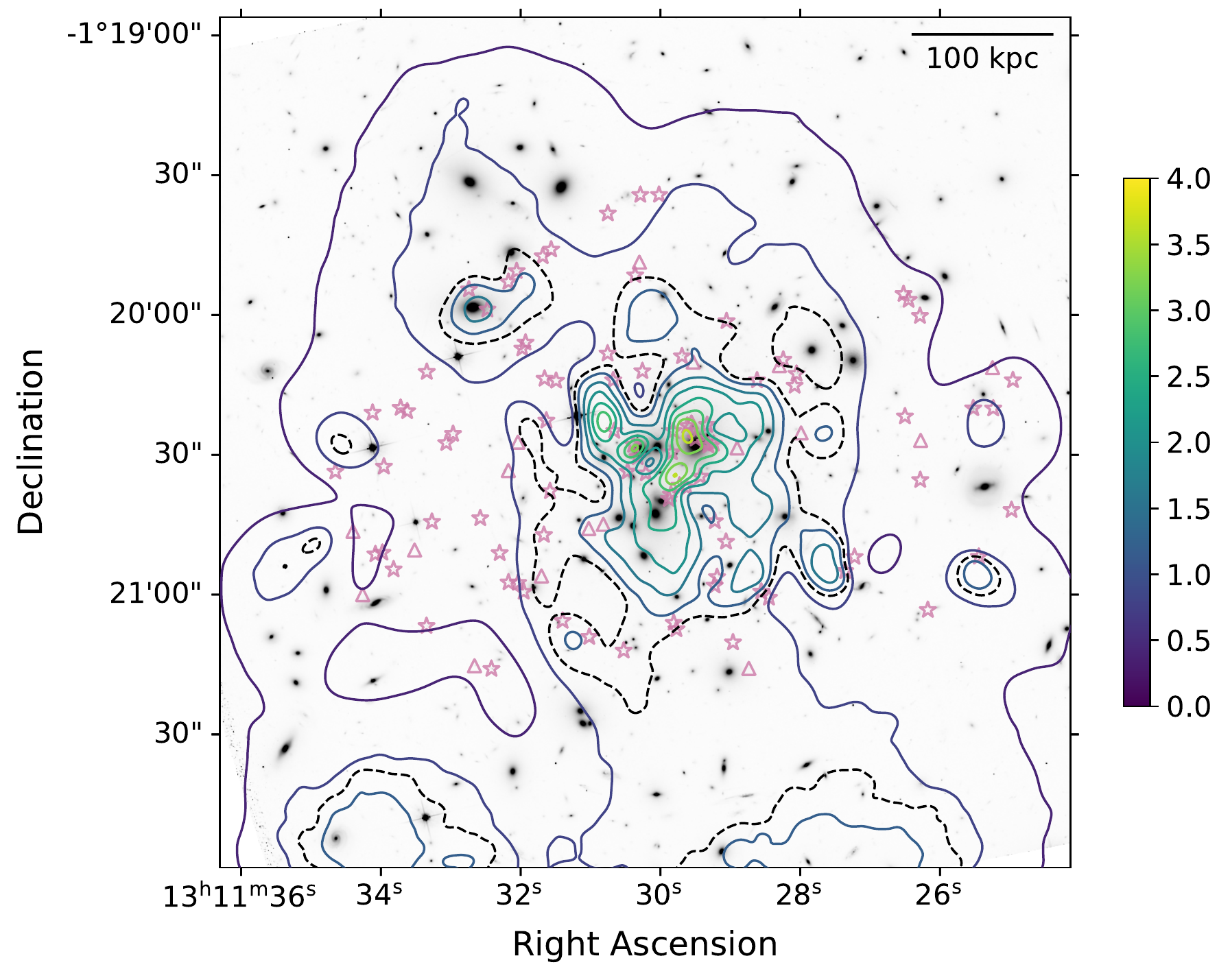}
         \caption{S-reconstruction}
     \end{subfigure}
     \caption{Same as Figure~\ref{fig:massmaps} but for low resolution reconstructions.}
    \label{fig:massmaps_lores}
\end{figure*}

\begin{figure*}
     \centering

     
     \begin{subfigure}[b]{0.475\textwidth}
         \centering
         \includegraphics[width=\textwidth]{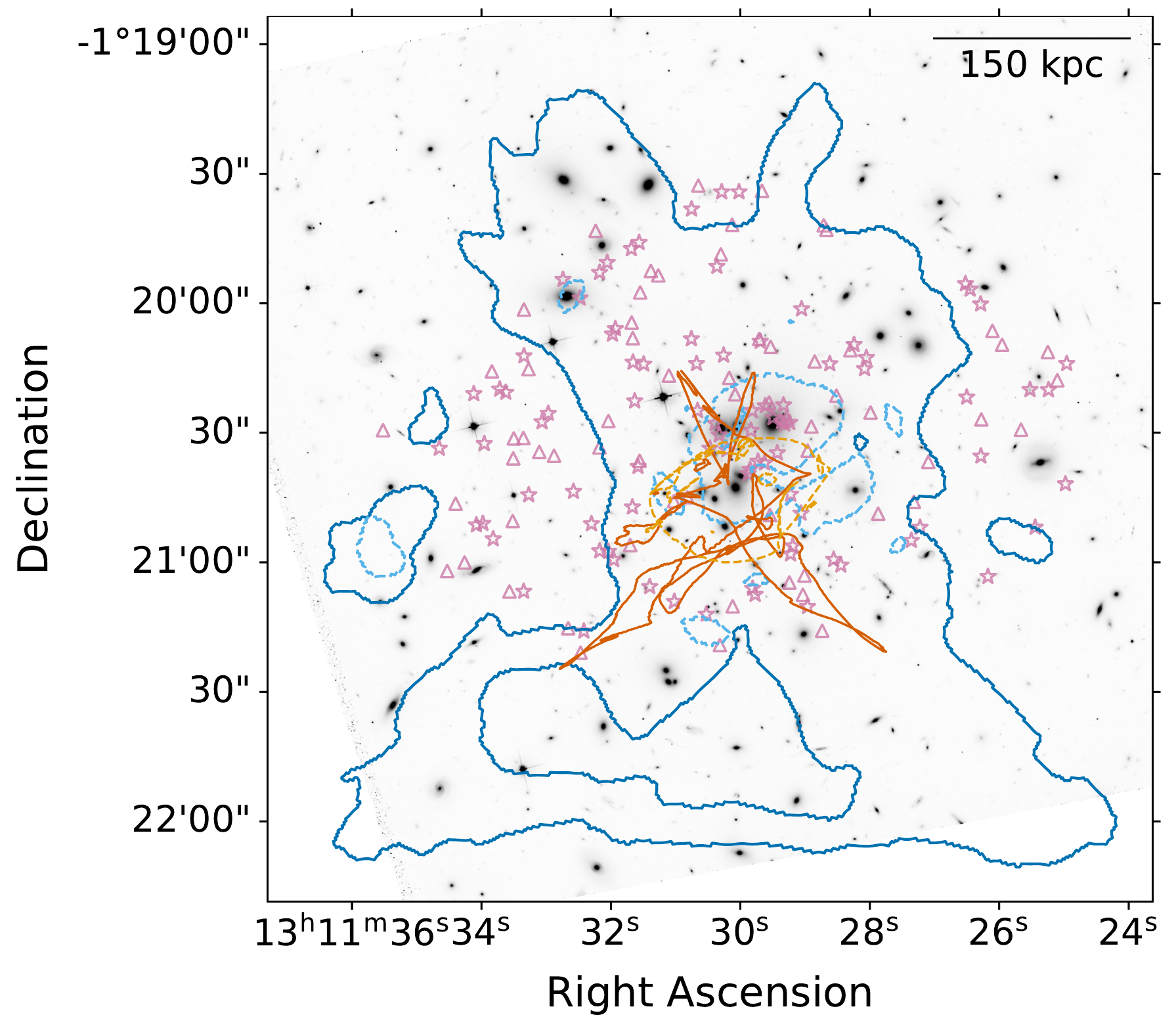}
         \caption{ALL-reconstruction}
     \end{subfigure}
     \quad
     \begin{subfigure}[b]{0.475\textwidth}
         \centering
         \includegraphics[width=\textwidth]{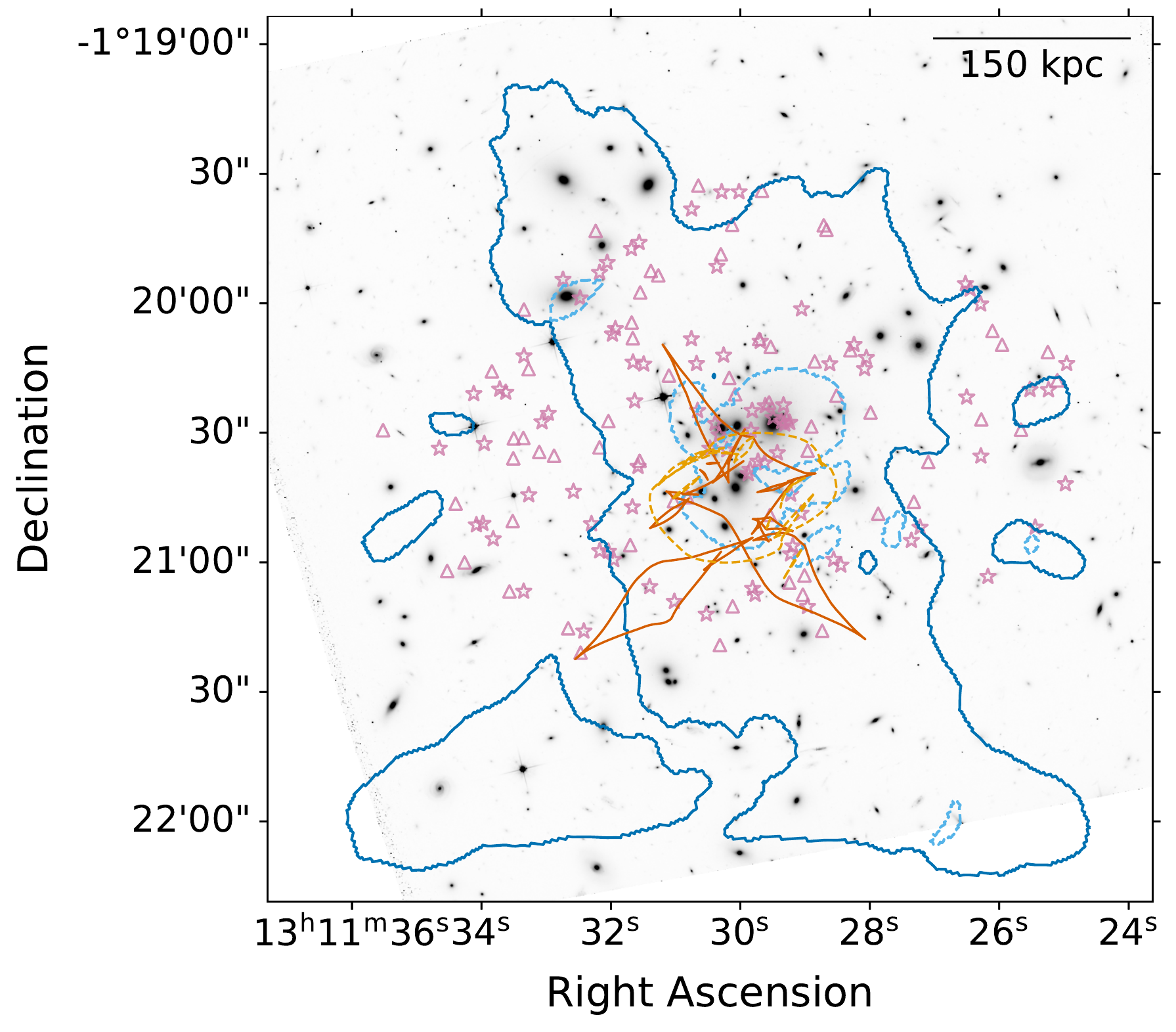}
         \caption{S-reconstruction}
     \end{subfigure}
     
        \caption{Same as Figure~\ref{fig:magmaps} but for low resolution reconstructions.}
        \label{fig:magmaps_lores}
\end{figure*}

\begin{figure*}
	\includegraphics[width=\textwidth]{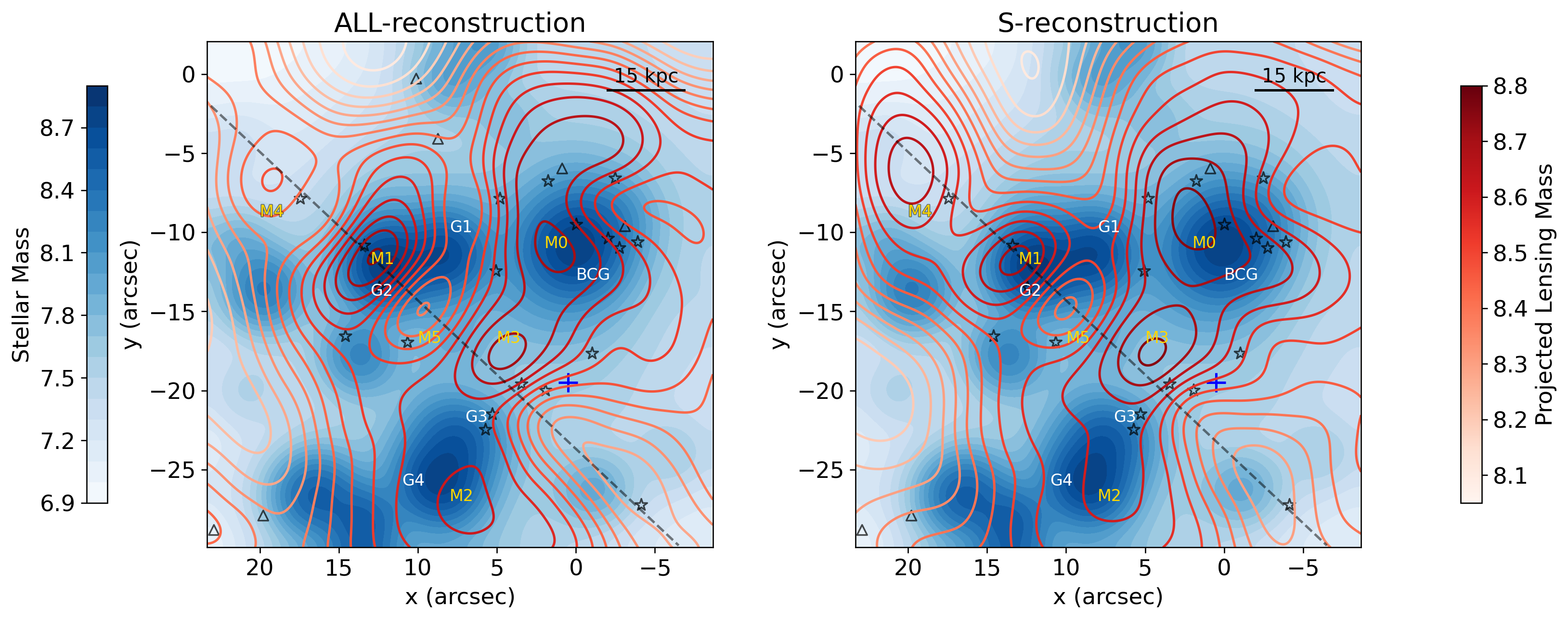}
    \caption{Same as Figure~\ref{fig:mm_hires} but for low resolution reconstructions.}
    \label{fig:mm}
\end{figure*}

\begin{figure}
	\includegraphics[width=\columnwidth]{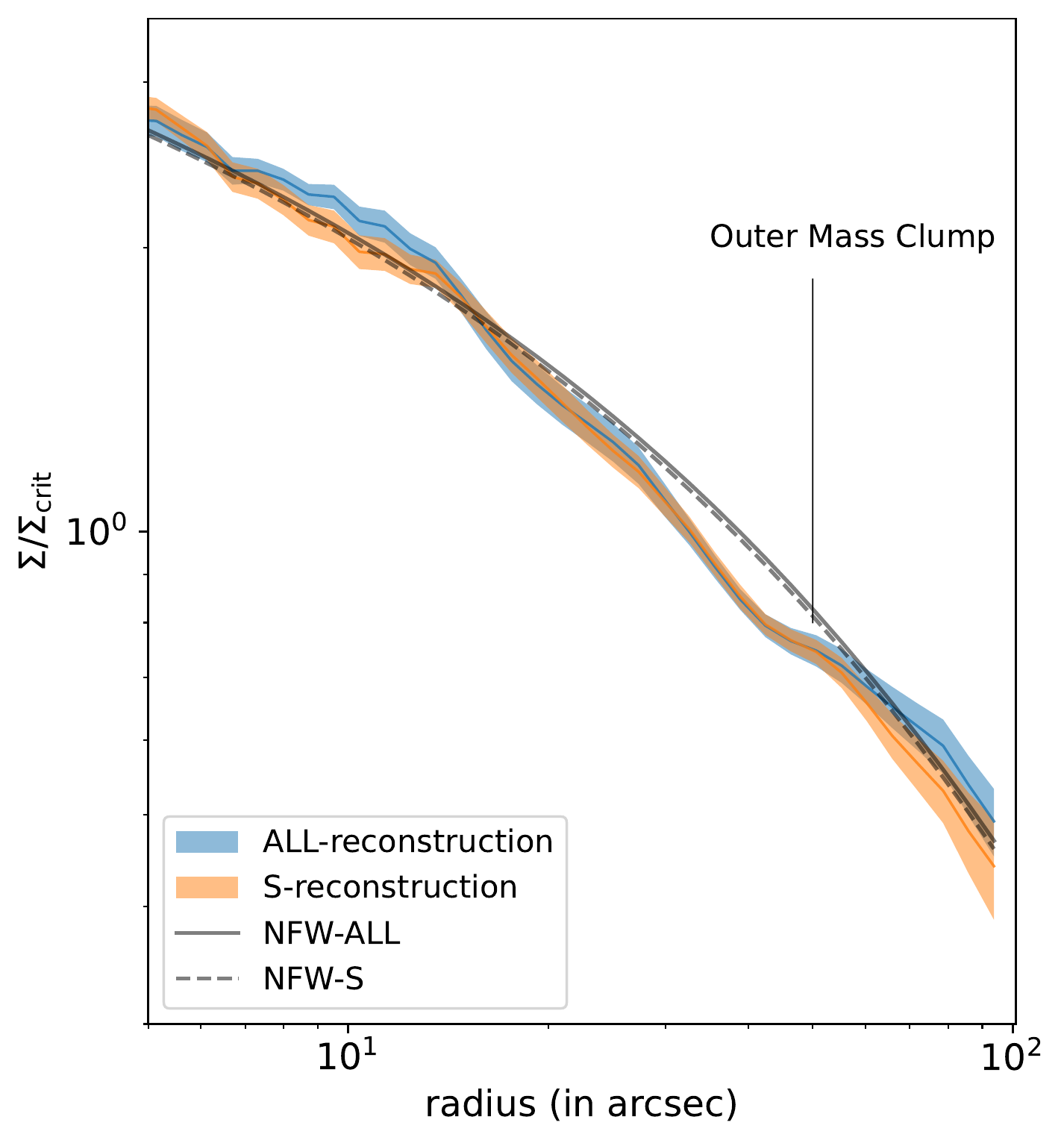}
    \caption{Same as Figure~\ref{fig:massprofile_allres} but for low resolution reconstructions.}
    \label{fig:massprofile}
\end{figure}


\begin{figure}
	\includegraphics[width=\columnwidth]{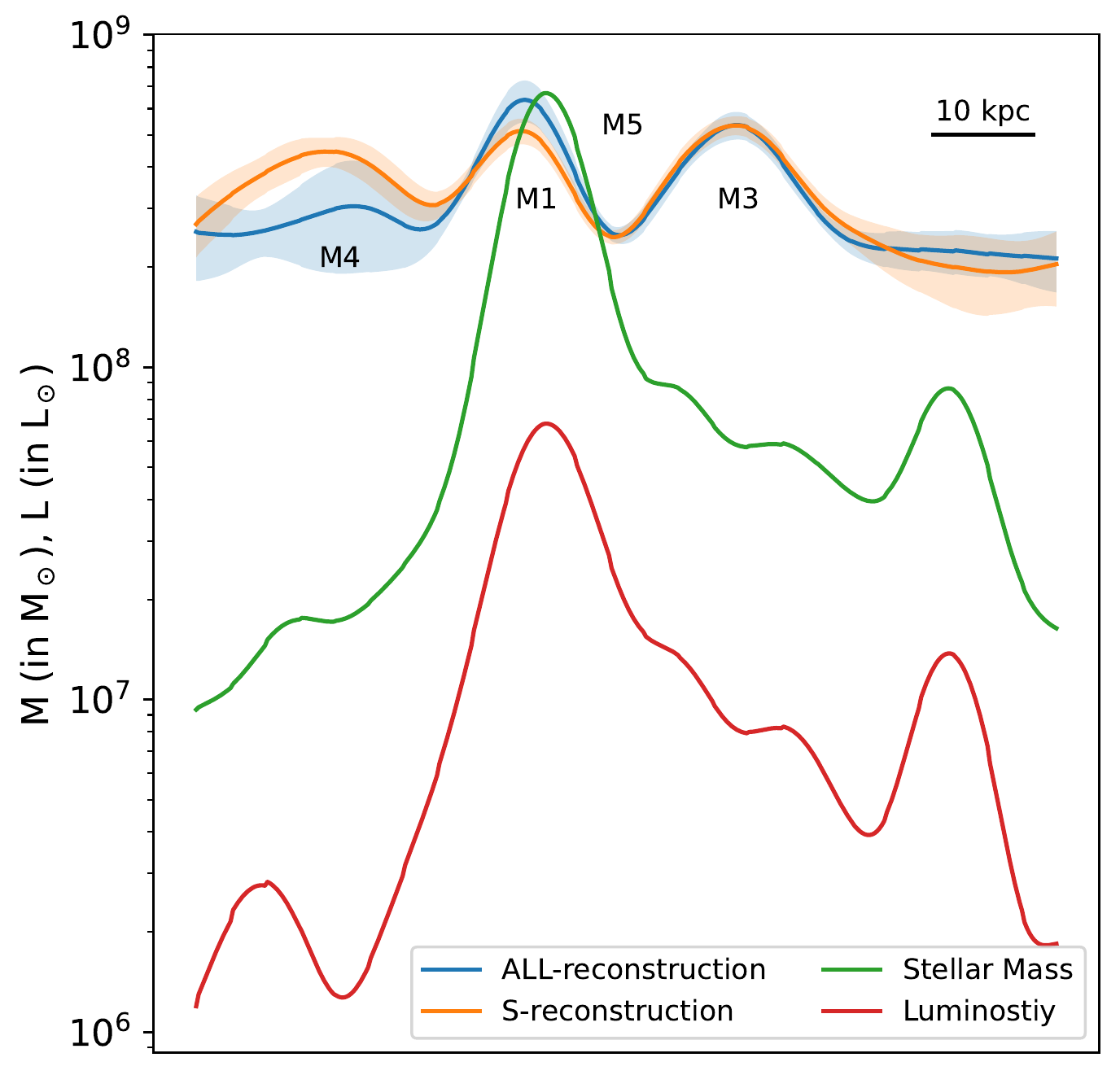}
    \caption{Same as Figure~\ref{fig:masscrossprofile_hires} but for low resolution reconstructions.}
    \label{fig:masscrossprofile}
\end{figure} 

\begin{figure*}
	\includegraphics[width=\columnwidth]{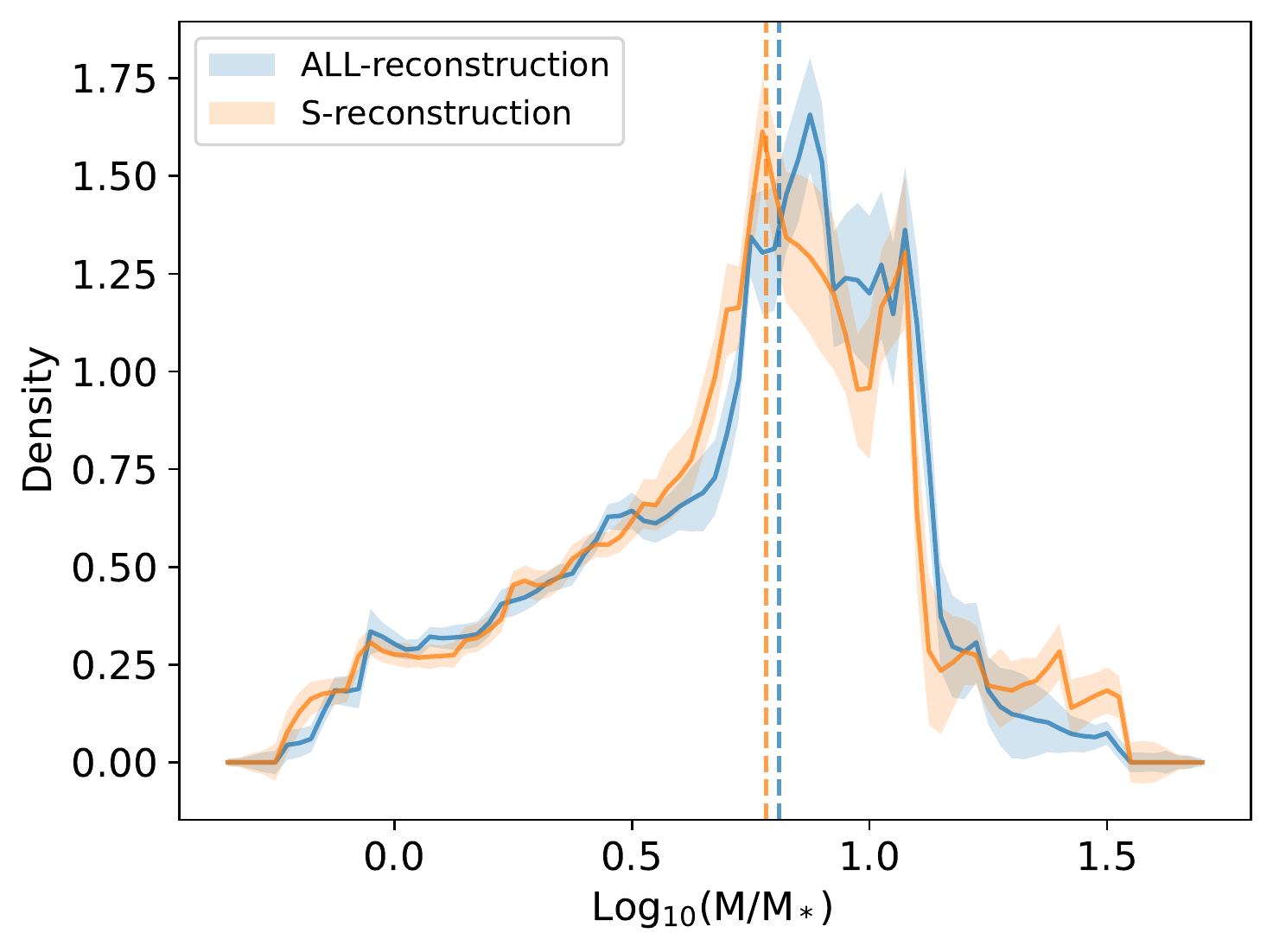}
     \includegraphics[width=\columnwidth]{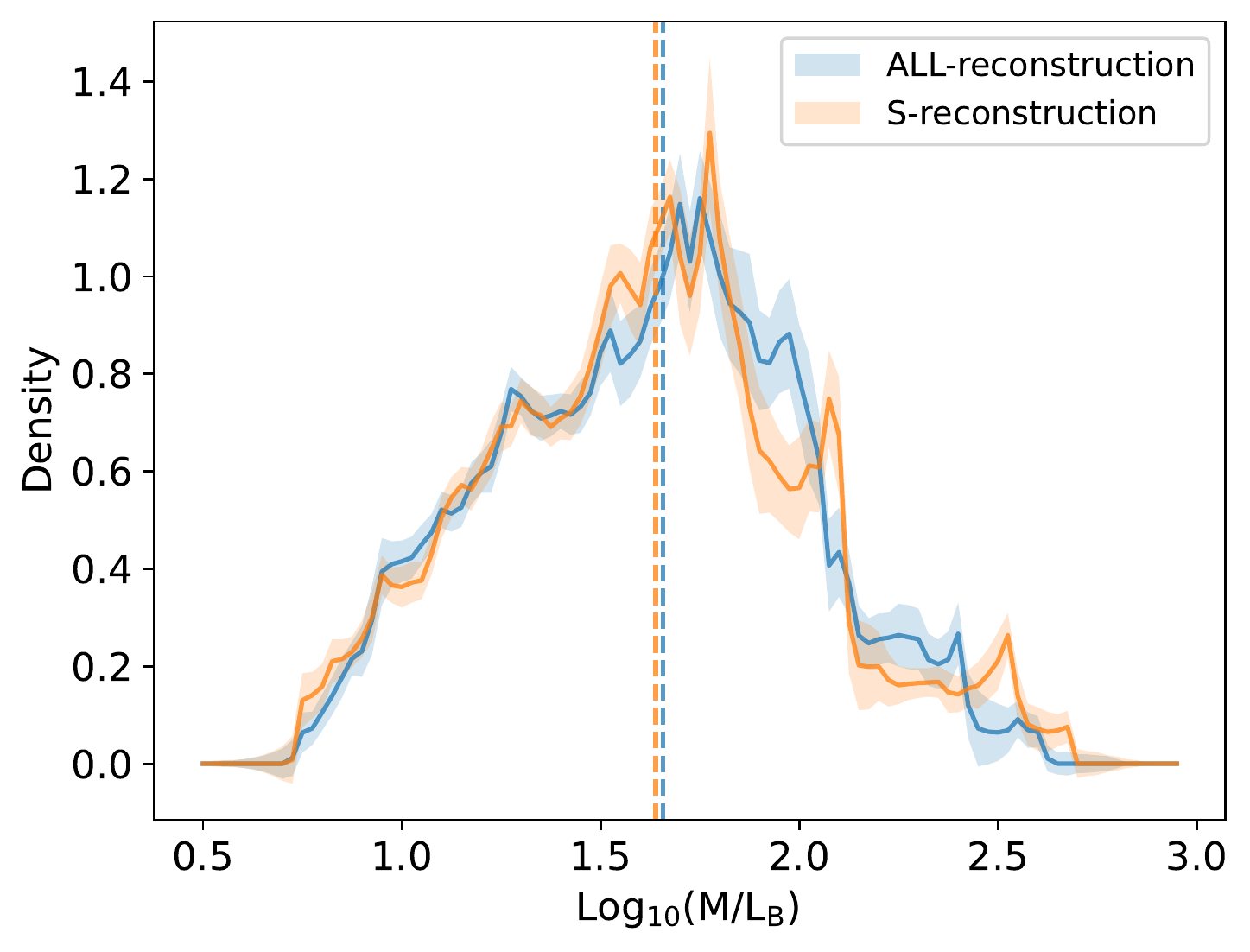}
    \caption{Same as Figure~\ref{fig:mmhist_hires} but for low resolution reconstructions.}
    \label{fig:mmhist}
\end{figure*}


\Review{In this Appendix we present reconstructions using lens plane \grale\ resolution typical of previous papers, which is lower than in the main paper by typically a factor of 0.5. \grale\ uses an adaptive grid, so the actual resolution depends on the position within the lens plane. Figures~\ref{fig:massmaps_lores}-\ref{fig:mmhist} show our main results. }

\Review{The mass peak nearest to the BCG is somewhat more offset here than with higher resolution reconstructions (see Figure~\ref{fig:mm}).
The average offset as recovered by the 40 different \textsc{grale} runs are $7.90\pm4.48$ kpc for the ALL-reconstruction, and $9.06\pm3.22$ kpc for the S-reconstruction. The offset in the average map is similar, and is  4.22 kpc and 7.55 kpc for the ALL- and S-reconstructions. These values are within offsets reported in \cite{Lauer2014}.}

Low resolution reconstructions show a lensing mass peak, M4, not associated with any light peaks, and not see in the high resolution reconstructions. However, due to lack of multiple images in its surrounding region it is not as well constrained as other lensing mass peaks. Additionally, there is a bright star to the left of that region making it harder to find multiple images or even a faint galaxy in that area. The uncertainty for this peak is high, especially for the ALL-reconstruction (see Figure~\ref{fig:masscrossprofile}), so its significance is low.

\Review{The dark mass peak M3 is prominent in these reconstructions as it is in the reconstruction presented in the main paper; see Figures~\ref{fig:mm} and \ref{fig:masscrossprofile}, which increases our confidence that this feature is real.}

\section{Multiple Image Systems for A1689}
\label{sec:imglist}

The list of multiple images is provided in Table~\ref{tab:imgdata}. More details on the lensing data can be found in Section~\ref{sec:lensingdata}.

\begin{figure*}
    \centering
	\includegraphics[width=0.9\textwidth]{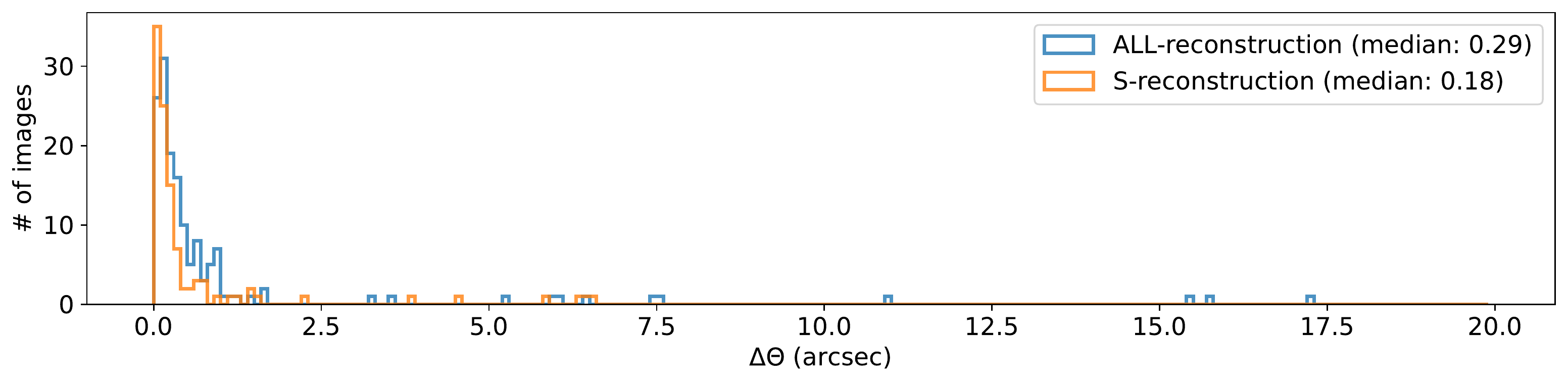}
    \caption{Histograms of lens plane separation values between the input multiple images and \textsc{grale} reconstructed images for high resolution reconstructions presented in the main paper. Median values are in the legend. \Review{All the images with $\Delta\theta> 7$ arcsec lie near the boundary of the lens plane image distribution, where constraints are sparse. High resolution S-reconstruction does not have any image with $\Delta\theta> 7$ arcsec.}}
    \label{fig:lprmshist}
\end{figure*}

\begin{figure*}
    \centering
	\includegraphics[width=0.9\textwidth]{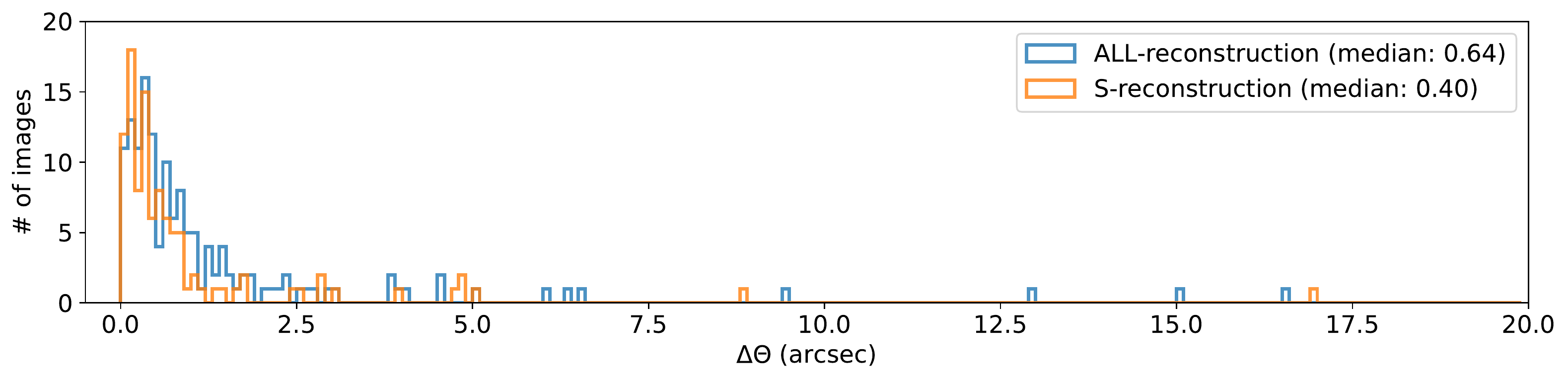}
    \caption{Similar to Figure~\ref{fig:lprmshist} but for low resolution reconstruction.}
    \label{fig:lprmshist_lowres}
\end{figure*}

\begin{figure*}
    \centering
	\includegraphics[width=0.8\textwidth]{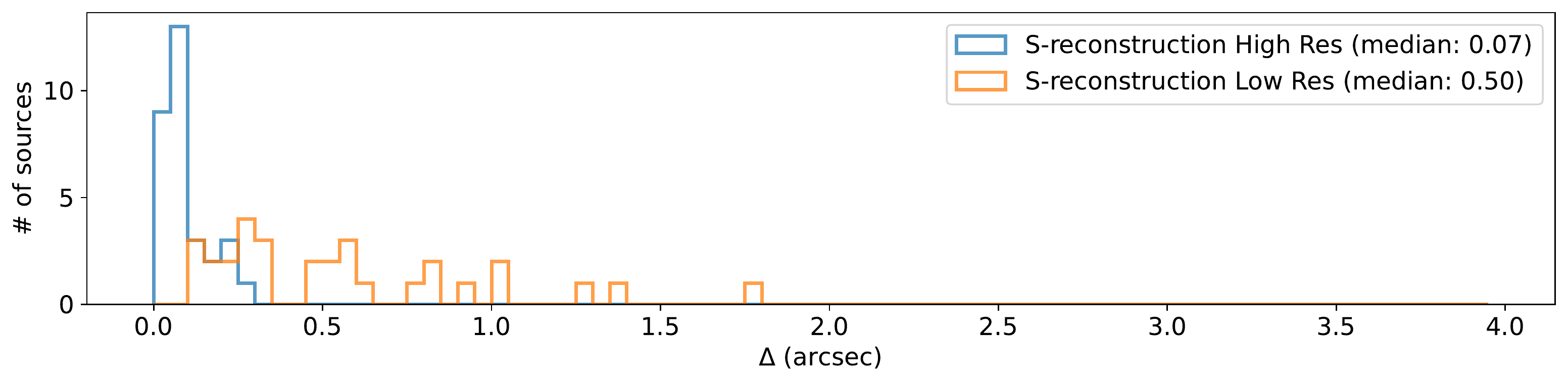}
    \caption{Histograms of source plane rms of backprojected images for S-reconstructions for both high resolution and low resolution cases. The high resolution reconstruction shows a much better source plane overlap, which is reflected in the median value.}
    \label{fig:sprmshist}
\end{figure*}

As a measure of how well \textsc{grale} uses and retrieves the input lensed images with the reconstruction process, one can compute the separations $(\Delta \Vec{\theta})$ between the observed lensed image positions and reconstructed image positions in the lens plane. Reconstructed lensed image positions are obtained by, first, back projecting the original images to the source plane and computing the average source positions for each system of multiple images using the reconstructed deflection angle values. This average back-projected location can then be forward lensed through the reconstructed lens mass distribution to obtain the reconstructed lensed image positions. Smaller lens plane separations between the original image positions and reconstructed image positions would imply a better reconstruction of the system. 

Sometimes the distances for all input images are combined into one single statistic as root mean square sum, namely the lens plane RMS \citep[e.g.,][]{williams2018,Ghosh2020}. Mathematically this can be expressed as, 
\begin{equation}
    \Delta \Vec{\theta}_{\rm rms}^2=\frac{1}{\sum_i J_i}\sum_{i=1,I}\left\{ \sum_{j=1,J_i} |\Vec{\theta}_{\text{original};i,j}-\Vec{\theta}_{\text{reconstructed};i,j}|^2 \right\},
    \label{eq:lprms}
\end{equation}
where $I$ is the number of systems with multiplicities (number of multiple images in each system) $J_i$ for the i-th system.

In Figure~\ref{fig:lprmshist} (Figure~\ref{fig:lprmshist_lowres}), we present a histogram of lens plane separation values for the input lensed images used for both of our reconstructions of the high (low) resolution runs. 
For some of the systems a few of the images have very large lens plane separation values skewing the distribution to the right. \Review{Images with $\Delta\theta>7$ arcsec are absent in S-reconstruction of the high resolution models. In other reconstructions, images with $\Delta\theta>7$ arcsec 
all lie near the boundary of the image region in the lens plane where the lensing constraints are sparse. } The number of these cases are higher for the ALL-reconstruction since it contains a larger number of less secure systems, and for low resolution runs. 
Since the lens plane rms values are sensitive to outliers, we can instead calculate the median of the lens plane separations, which are found to be 0.40 arcsec and 0.64 arcsec for the S- and ALL- reconstructions. This shows a small improvement of the input image utilization by \textsc{grale} when only the secure systems are used as input.

\begin{table*}
\centering
\caption{Details of the multiple lensed images for A1689 retrieved from the existing literature. Full table will be available online. [B05: \citet{Broadhurst2005}, L07: \citet{Limousin2007}, C10: \citet{Coe2010}, D15: \citet{Diego2015}, B16: \citet{Bina2016}, A16: \citet{Alavi2016}]}  
\label{tab:imgdata}
\begin{minipage}{\textwidth}
\renewcommand{\thempfootnote}{\fnsymbol{mpfootnote}}
\begin{center}

\begin{tabular}{cccccccc}
\hline
    ID & R.A.       & Decl.       &  Redshift  & Redshift Type  & Reference        \\
     & (deg.)       & (deg.)       &   (z)  &    &        \\ \hline
    1.1  & 197.860217 & -1.332431 & 3.038  & spectroscopic & B05    \\
    1.2  & 197.859538 & -1.333386 & 3.038  & spectroscopic & B05     \\
    1.3  & 197.874054 & -1.352064 & 3.038  & spectroscopic\footnote[2]{\label{note1}Spectroscopic redshift is confirmed by MUSE \citep{Bina2016}.} & B05    \\
    1.4  & 197.887775 & -1.340964 & 3.038  & spectroscopic & B05    \\
    1.5  & 197.883050 & -1.334975 & 3.038  & spectroscopic & B05    \\
    1.6  & 197.874383 & -1.344028 & 3.038  & spectroscopic\footref{note1} & B05    \\ 
    2.1  & 197.860517 & -1.332081 & 2.53   & spectroscopic & B05    \\
    2.2  & 197.887371 & -1.340419 & 2.53   & spectroscopic & B05    \\
    2.3  & 197.883242 & -1.335325 & 2.53   & spectroscopic & B05    \\
    2.4  & 197.874217 & -1.351681 & 2.53   & spectroscopic & B05    \\
    2.5  & 197.874504 & -1.344300 & 2.53   & spectroscopic & B05    \\
    $\cdots$ \\
    \hline
\end{tabular}
\end{center}
\end{minipage}
\end{table*}





\bsp	
\label{lastpage}

\end{document}